\documentclass[twocolumn,           
               showpacs,            
               preprintnumbers,     
               aps,                 
               prd,                 
               a4paper,             
               superscriptaddress,  
               nofootinbib,         
               tightenlines,        
               floats,floatfix      
               ]{revtex4}

\usepackage{graphicx}
\usepackage{subfigure}
\usepackage{latexsym}
\usepackage{amsmath,amssymb}        
\usepackage[draft=false]{hyperref}
\usepackage{threeparttable}

\newcommand{\de}{\mathrm{d}}

\newcommand{\reffig}[1]{Fig.~\ref{#1}}
\newcommand{\refeq}[1]{Eq.~(\ref{#1})}
\newcommand{\reftab}[1]{Table~\ref{#1}}

\input colordvi

\begin{document}



\title{Direct reconstruction of the quintessence potential}
\author{Martin Sahl\'en}
\affiliation{Astronomy Centre, University of Sussex, Brighton BN1 9QH, United 
Kingdom}
\affiliation{Department of Physics, Stockholm University, AlbaNova University 
Centre, SE-106 91 Stockholm, Sweden}
\author{Andrew R.~Liddle}
\affiliation{Astronomy Centre, University of Sussex, Brighton BN1 9QH, United 
Kingdom}
\author{David Parkinson}
\affiliation{Astronomy Centre, University of Sussex, Brighton BN1 9QH, United 
Kingdom}
\date{\today}
\pacs{98.80.-k \hfill astro-ph/0506696}
\preprint{astro-ph/0506696}


\begin{abstract}
We describe an algorithm which directly determines the quintessence
potential from observational data, without using an equation of state
parametrisation. The strategy is to numerically determine
observational quantities as a function of the expansion coefficients
of the quintessence potential, which are then constrained using a
likelihood approach. We further impose a model selection criterion,
the Bayesian Information Criterion, to determine the appropriate level
of the potential expansion. In addition to the potential parameters,
the present-day quintessence field velocity is kept as a free
parameter. Our investigation contains unusual model types, including a
scalar field moving on a flat potential, or in an uphill direction,
and is general enough to permit oscillating quintessence field
models. We apply our method to the `gold' Type Ia supernovae sample of
Riess {\it et al.}~(2004), confirming the pure cosmological constant model
as the best description of current supernovae luminosity--redshift
data.  Our method is optimal for extracting quintessence parameters
from future data.
\end{abstract}

\maketitle


\section{Introduction}

Quintessence, a scalar field slowly rolling on its potential, remains
one of the most attractive possibilities for explaining the observed
acceleration of the Universe (for reviews, see Ref.~\cite{reviews}). A
key goal for future observational programs is to seek definitive
evidence for variation in the dark energy density with redshift, which
would exclude a cosmological constant. In that event, one would then
seek an optimal determination of dark energy properties in the hope of
relating them to fundamental physics.

In this paper we assume from the outset that single-field quintessence
remains a viable description of observational data, i.e.~that it has
successfully passed tests against other dark energy paradigms. Our aim
is then to obtain optimal constraints on the quintessence
potential. We do this by passing directly between the quintessence
potential and the observable quantities, focusing in this paper on the
luminosity--redshift relation of type Ia supernovae (SNIa). We
parametrise the potential, and then constrain those parameters, along
with global properties of the Universe, via a likelihood
analysis. Additionally, we use model selection criteria in order to
select the preferred level of parametrisation of the potential.

Although a variant on the general scheme of reconstruction, our
approach is distinct from those already in the literature \cite{rec}
in that we do not rely on a parametrisation of the dark energy
equation of state, which then must be related to the dark energy
potential via relations which may be approximate (see Guo {\it et 
al.}~\cite{GOZ} for relations in some particular cases). The work closest in
spirit to our own is that of Simon {\it et al.}~\cite{simon}, who consider
an extremely general action and expand the quintessence potential in
Chebyshev polynomials (in the redshift range of available data). They
relate the expansion coefficients to the redshift evolution of the
matter density and Hubble parameter. Those quantities are then
extracted from observations and processed into constraints on the
potential, those constraints however being as a function of redshift
rather than scalar field value. Their treatment is roughly analogous
to the inflationary reconstruction method whereby observables such as
the spectral index and tensor amplitude are obtained from data, and
then related to the inflationary potential via the slow-roll
approximation \cite{CKLL}. Our present paper is analogous to the
direct inflaton potential reconstruction method proposed by Grivell
and Liddle \cite{GL}, where the observed power spectra are predicted
numerically directly from the inflation potential.

\section{Formalism}

Our set-up is relatively straightforward. We assume that the
quintessence field $\phi$ has a potential $V(\phi)$, which we expand
as a power series
\begin{equation}
V(\phi) = V_0 + V_1 \phi + V_2 \phi^2 + V_3 \phi^3 + 
\cdots \,,
\end{equation}
where the field is measured in Planck units and (without loss of
generality) we take $\phi$ to be presently zero.  Note that when we
fit these parameters, even in the case of ``complete'' data we do not
necessarily obtain the MacLaurin expansion of the true potential to
the same order as this is generally not the best polynomial fit over
an interval (in the least-squares or minimax sense). We choose not to
use a Chebyshev series as in Ref.~\cite{simon} for the following
reasons. Firstly, when fitting, the coefficients of a Chebyshev series
would just be linear combinations of the coefficients of a monomial
basis polynomial of the same order, so the fits are the
same. Furthermore, Chebyshev polynomials would depend on the range
$\phi$ takes which in turn depends on the model parameters. Lastly,
because we do not fit the potential expansion directly, but rather
through a function depending on an integral of the potential,
expanding the potential in orthogonal polynomials will not guarantee
uncorrelated coefficients.

The scalar field obeys the equation
\begin{equation}
\ddot{\phi} + 3 H \dot{\phi} = - \frac{dV}{d\phi} \,,
\end{equation}
with the Hubble parameter given by the Friedmann equation
\begin{equation}
H^2 = \frac{8\pi G}{3} \left(\rho_{{\rm m}} + \rho_{\rm \phi} \right) \,.
\end{equation}
Here $\rho_{{\rm m}}$ is the matter density and $\rho_{\rm \phi} =
\dot{\phi}^2/2 + V(\phi)$ the quintessence density. We assume spatial
flatness throughout, though the generalisation to the non-flat case
would be straightforward.  Since then $\Omega_{\rm m} + \Omega_{\phi}
= 1$ we have the initial condition
\begin{equation}
\label{eq:phidot}
\dot{\phi}_{\rm 0} = \pm \sqrt{2\left[(1-\Omega_{\rm m})\rho_{\rm c,0} - V_0 
\right]} \,.
\end{equation}
We allow $\dot{\phi}_0$ to take either sign but results are symmetric
under simultaneous reversal of its sign and of odd-order expansion
coefficients.

In this article we focus on SNIa data, and hence the observational
quantity we need to predict is the luminosity distance as a function
of redshift.  The luminosity distance is given by
\begin{equation}
\label{eq:lumdistz}
d_{\rm L}(z; \mathbf{\Theta}) = \frac{{\cal D}_{\rm L}(z; 
\mathbf{\Theta})}{H_{\rm 0}}  
\,,
\end{equation}
where
\begin{equation}
\frac{{\cal D}_{\rm L}(z; \mathbf{\Theta})}{1+z} =
\int_0^z  \frac{{\rm d}z'}{\left[ \Omega_{\rm m}(1+z')^3 +
( 1-\Omega_{\rm m}) e^{F(z'; \mathbf{\Theta})} \right]^{1/2} } \,
\end{equation}
is the Hubble-constant-free luminosity distance (for the low redshifts
we consider there is no contribution from radiation), and
\begin{equation}
F(z; \mathbf{\Theta}) =  3 \int_0^{z} \left(1+w_{\rm \phi}(z'; 
\mathbf{\Theta})\right) 
{\rm d}\ln(1+z') \,.
\end{equation}
$\mathbf{\Theta}$ is the parameter vector describing the model, and
\begin{equation}
w_{\rm
\phi} \equiv \frac{p_{\rm \phi}}{\rho_{\rm \phi}} = \frac{\dot{\phi}^2/2 - 
V(\phi)}{\dot{\phi}^2/2 + V(\phi)} \,,
\end{equation}
is the quintessence equation of state.  The apparent magnitude $m(z;
\mathbf{\Theta})$ of a type Ia supernova can be expressed as
\begin{equation}
m(z; \mathbf{\Theta}) = M + 5\log_{10} \left( \frac{d_{\rm L}(z; 
\mathbf{\Theta})}{\rm 
Mpc} \right) + 25 \,,
\end{equation}
where $M$ is the absolute magnitude of SNIa (supposing they are
standard candles). The distance modulus $\mu(z; \mathbf{\Theta})$ is
defined as
\begin{equation}
\mu(z; \mathbf{\Theta}) \equiv m(z; \mathbf{\Theta}) - M \,.
\end{equation}

In the following, we should in principle keep $M$ as a free
parameter. To this end, we define our ``observational'' quantity to be
\begin{equation}
\mu_i \equiv m_i - M \,,
\end{equation}
where $m_i$ is a measurement of $m(z_i; \mathbf{\Theta}_{\rm true})$
(with $\mathbf{\Theta}_{\rm true}$ the projection of the parameters of
the ``true'' model of the Universe onto our model and its parameter
space).  Supernovae observations measure $m$, but typically report a
distance modulus
\begin{equation}
\mu_i^* \equiv m_i - M^* \,,
\end{equation}
where $M^*$ is some estimate of the absolute magnitude.  The relevant
quantity for fits is thus
\begin{equation}
\mu_i - \mu(z_i; \mathbf{\Theta}) = \mu_i^* + \eta - 5\log_{10} {\cal
D}_{\rm L}(z_i; \mathbf{\Theta}) \,,
\end{equation}
where
\begin{equation}
\eta \equiv 5\log_{10} \left(H_{\rm 0}\,\,{\rm Mpc} \right)+ \Delta M - 25
\end{equation}
and $\Delta M \equiv M^* - M$.  Because of the perfect degeneracy
between $M$ and $\log_{10} H_0$, and the fact that the equations are
otherwise independent of these parameters, our effective
$D$-dimensional parameter vector is
\begin{equation}
\mathbf{\Theta} = (\eta, \dot{\phi}_{\rm 0}, V_0, \dots, V_{D-3}) \,.
\end{equation}
Recall that $\dot{\phi}_{\rm 0}$ and $V_0$ determine the matter density 
through \refeq{eq:phidot}.

To end this section, we note that our formalism includes some
possibilities which are not commonly considered. Even if the potential
is truncated as a constant, the present field velocity remains a free
parameter and so the scalar field can move on this flat
potential.\footnote{This situation is equivalent to having a pure
cosmological constant plus a stiff fluid with $w\equiv1$.} To regain
the cosmological constant case we must make the additional assumption
that this velocity is zero. Further, the field may be rolling uphill;
this may seem unlikely but is valid phenomenologically and might occur
in models where the field has recently passed beyond a minimum. Our
analysis can also generate models where the scalar field has undergone
one or more oscillations about a minimum in the recent past.

\section{Data analysis}

\subsection{Likelihood analysis}

We carry out a likelihood analysis of the models in comparison to the
observational data from Riess {\it et al.}~\cite{Riess:2004nr}. We
use the 157 SNIa of the `gold' sample.

With a prior distribution $P(\mathbf{\Theta})$, the posterior
probability of the parameters $\mathbf{\Theta}$, given the data set,
is given according to Bayes' theorem by
\begin{equation}
P(\mathbf{\Theta} | {\rm data}) \equiv \frac{1}{{\cal Z}} 
{\cal L}({\rm data}|\mathbf{\Theta}) P(\mathbf{\Theta})
= \frac{1}{{\cal Z}} e^{-\chi^2(\mathbf{\Theta})/2} P(\mathbf{\Theta}) \,,
\end{equation}
where
\begin{equation}
\chi^2(\mathbf{\Theta}) = \sum_{i=1}^{N} \frac{\left(\mu_i - \mu(z_i; 
\mathbf{\Theta})\right)^2}{\sigma_i^2}
\end{equation}
is summed over all $N$ data points, and ${\cal Z}~=~\int {\cal L}({\rm
data}|\mathbf{\Theta}) P(\mathbf{\Theta}) {\rm d}\mathbf{\Theta}$ is a
normalisation constant, irrelevant for parameter fitting (but as it is
the Bayesian evidence, highly relevant for model selection as
discussed below).  Here $\mu_i$ and $\sigma_i$ are the observed
distance moduli and their standard deviations, $z_i$ the redshift of
the observed supernova and $\mu(z_i; \mathbf{\Theta})$ the distance
modulus predicted for the redshift $z_i$ by our model with parameters
$\mathbf{\Theta}$.

To estimate parameters we wish to find $P(\mathbf{\Theta}|{\rm data})$
explicitly as a function of $\mathbf{\Theta}$.  This is in general
non-trivial, and the standard approach is to explore the parameter
space in some way and keep a histogram characterizing
$P(\mathbf{\Theta}|{\rm data})$.  We choose to explore the parameter
space using an MCMC approach
\cite{Gilks,Lewis:2002ah,Verde:2003ey,Dunkley:2004sv}. MCMC
calculations are generally preferable over grid methods as they scale
approximately linearly with the dimension of the problem, rather than
exponentially.

Our MCMC algorithm is the following, and makes use of relatively
standard step optimisation and convergence/mixing testing.

\vskip 4pt
\noindent
1. The starting points for the Markov chains are chosen to be close to
the expected high-likelihood region with some random spread, checking
that they satisfy the priors.

\vskip 4pt
\noindent
2. Starting with an initial best guess for the covariance matrix of
the underlying distribution, we optimise the step sizes of the
Gaussian trial distribution with the iteration rule \cite{Gilks}
\begin{equation}
\label{eqn:cov}
{\mathbf C_T}_i=(2.4^2/D){\mathbf C}_{i-1} \,,
\end{equation}
where ${\mathbf C_T}_i$ is the $i^{\rm th}$ estimate of the covariance
matrix of the trial distribution, $D$ is the number of parameters and
${\mathbf C}_{i-1}$ the covariance matrix of the $(i-1)^{\rm th}$
chain produced (with ${\mathbf C}_{0}$ our initial best guess). We use
chains of 10 000 elements for the optimisation process, and continue
updating the trial distribution until there is no significant increase
in the sampling efficiency (assessed by comparing the eigenvalues of
the covariance matrices). Between each iteration, the parameter space
is rotated to the eigenspace of the new covariance matrix, to maximise
the efficiency in exploring the shape of the likelihood distribution.

\vskip 4pt
\noindent
3. The full production run is started.  A set of $m$ chains with $n$
elements each is generated, and only these are used for the final
analysis.  We generate well separated starting points as before for
each of the chains.  The chains are tested for convergence and mixing
using the Gelman--Rubin test \cite{GelmanRubin,Gilks}, which compares
the variances within a chain to the variances between chains, which in
the asymptotic limit should give a Gelman--Rubin ratio $R = 1$. We
require $R < 1.05$ for each parameter.  A consistently high and
non-convergent Gelman--Rubin ratio is indicative of a very loosely
constrained parameter.

\vskip 4pt

In the above, all calculations of covariances and means are done by
first dropping an initial burn-in section from the chain. We define
the burn-in section following Tegmark {\it et
al.}~\cite{Tegmark:2003ud} as the elements in the chain from the
beginning up to the first element to have a likelihood value above the
median likelihood value of the whole chain.  The chains were analysed
using a slightly modified version of {\tt GetDist} provided with {\tt
CosmoMC} \cite{Lewis:2002ah} (again with burn-in sections excluded).

We impose two important constraints on the behaviour of the cosmology
\emph{within the redshift range} $0 \le z \le 2$, and hence as priors
on the parameters.  Firstly, the total energy density of the universe
must remain positive at all times to exclude collapsing epochs, and
secondly we need to avoid models where the kinetic energy would
dominate at early epochs (such domination is permitted by the SNIa
data alone, but is inconsistent with other data as discussed later).
We limit the kinetic contribution to $\Omega_{\rm kin} < 0.5$ for $z
\ge 1$ --- see Section~\ref{s:kin}.  Most marginalised posterior
likelihoods are fairly insensitive to the particular choice of upper
limit. However, some marginalised posteriors involving $\dot{\phi}_0$
do change significantly. This will be discussed further in the Results
section.  Additionally we naturally impose $0 \le \Omega_{\rm m} \le
1$.  No other priors (e.g. on $H_0$) were found to be necessary to
obtain acceptable cosmologies.

\subsection{Model selection}

The order of the power series of the potential can be freely chosen,
and the results obtained will obviously depend on the order to which
it is taken, with parameters becoming less and less constrained as the
order increases. In addition to a determination of the best-fitting
parameters within a given model, one therefore needs to compare the
different models (i.e.~expansions to different orders) in order to
determine which is the preferred fit to the data.

Since models with more parameters will always lead to an improved
best-fit model, one must use model selection statistics
\cite{jeff,mackay,Lid}. These set up a tension between the number of
model parameters and the goodness of fit. In the context of Bayesian
inference the best such statistic is the Bayesian evidence
\cite{jeff,mackay}; for an application to SNIa data see
Ref.~\cite{SWB}. The evidence is however difficult to calculate, and
in this paper we use a simpler statistic, the Bayesian Information
Criterion (BIC) \cite{Schwarz,Lid}, which gives a crude approximation
to the evidence. The BIC is given by
\begin{equation}
\mathrm{BIC} = -2\ln{\cal L}_{{\rm max}} + D \ln N \,,
\end{equation}
where ${\cal L}_{{\rm max}}$ is the likelihood of the best-fitting
parameters for that model, $D$ the number of model parameters, and $N$
the number of datapoints used in the fit. Models are ranked with the
lowest value of the BIC indicating the preferred model. A difference
of 2 for the BIC is regarded as positive evidence, and of 6 or more as
strong evidence, against the model with the larger value
\cite{jeff,Muk98}.

It is worth mentioning that although we specifically consider a
quintessence scenario, a model selection result favouring more than
one potential parameter would indicate a dynamical dark energy
component more generally, since for every choice of $\{H(z),\rho_{{\rm
m}}(z)\}$ there exists a corresponding quintessence potential, by
virtue of Picard's existence theorem for ODE's (demonstrated
explicitly in e.g.~Ref.~\cite{Padmanabhan:2004av}).

\begin{figure}[t]
\includegraphics[width=\linewidth]{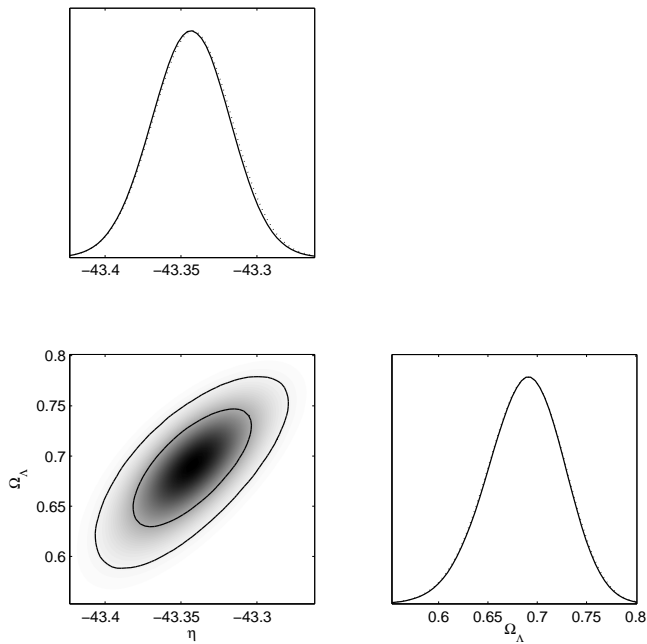}
\caption{One and two-dimensional likelihood distributions for
$D=2$. Solid lines are marginalised 1D likelihoods and dotted lines
mean 1D likelihoods. Solid 2D contours represent 68\% and 95\% regions
of the marginalised distribution, and shading reflects the mean
distribution.}
\label{fig:d2_tri}
\end{figure}

\section{Results}

The maximum likelihood value and parameters were estimated using the
approach described in the preceding Section. We investigated the
cosmological constant case and then cases of one, two and three
potential parameters (i.e. polynomial orders zero, one, two). Since
solving the necessary ODEs is not computationally intensive, we can
generate very long chains. For each scenario, 10 chains each
containing 1 000 000 elements were obtained.

\subsection{Parameter estimation}

\subsubsection{Cosmological constant ($D=2$)}

As a check, we investigate the case of a cosmological constant. Our
parameter vector is
\begin{equation}
\mathbf{\Theta} = (\eta, V_0) \,,
\end{equation}
since $\dot{\phi}_0 = 0$.  Indeed we obtain the well-known results for
SNIa data, as seen in \reffig{fig:d2_tri}.The constraint on the matter
density is shown along with that of other models in
\reffig{fig:omegam_combo}.

\subsubsection{Constant potential with kinetic energy ($D=3$)}

\begin{figure}[t]
\includegraphics[width=\linewidth]{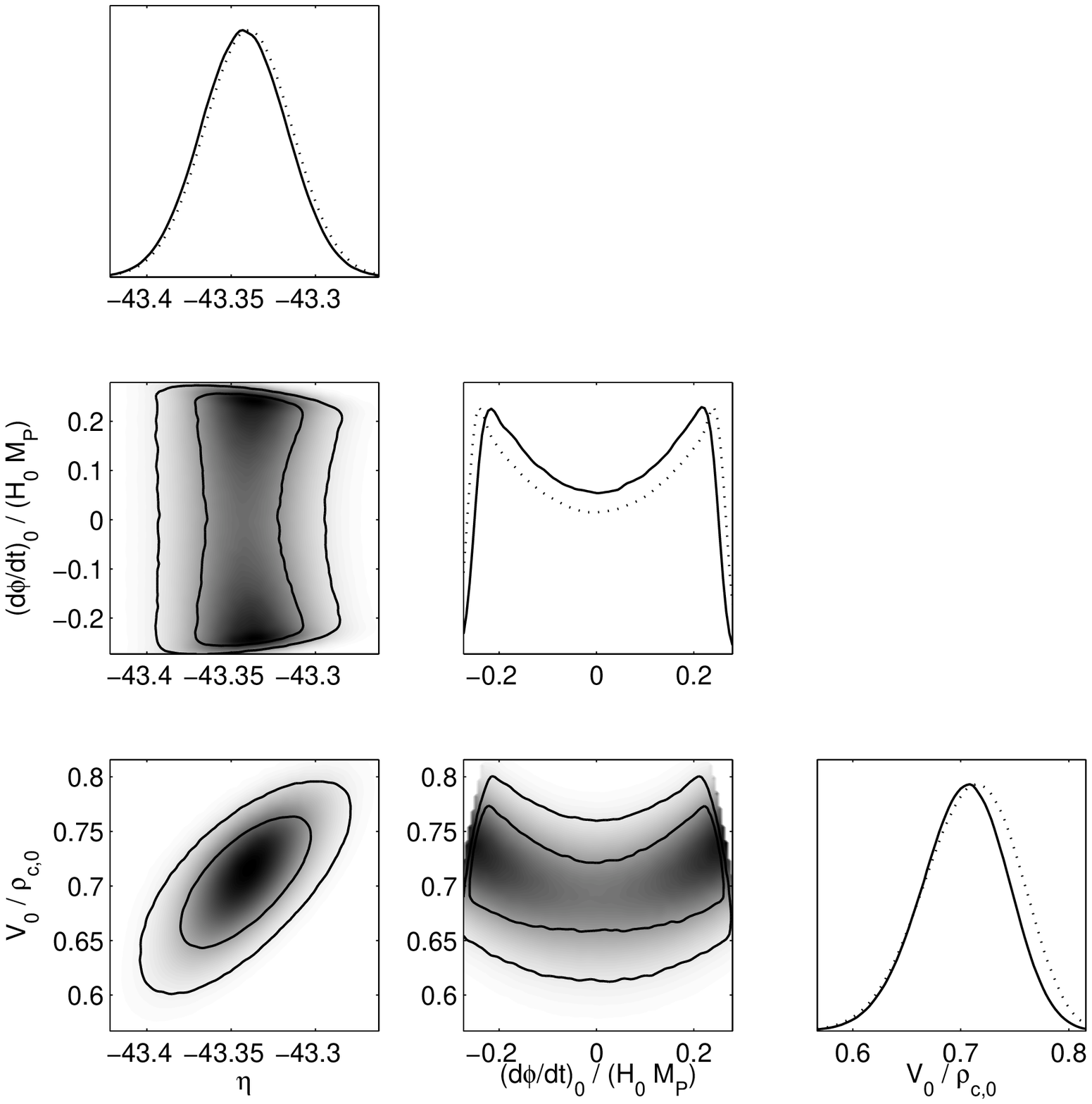}
\caption{As \reffig{fig:d2_tri} for $D=3$.}
\label{fig:d3_tri}
\end{figure}

Allowing a non-zero kinetic contribution on a constant potential means
our parameter vector is
\begin{equation}
\mathbf{\Theta} = (\eta, \dot{\phi}_0, V_0) \,.
\end{equation}
This does, unsurprisingly, improve the fit to data relative to the
pure cosmological constant. However, looking at the likelihood
distributions in \reffig{fig:d3_tri}, we clearly see that
$\dot{\phi}_0 = 0$ is not excluded at a statistically-significant
level. The bimodality in the $\dot{\phi}_0$ distributions is due to
the model depending only on $\dot{\phi}_0^2$.

\begin{figure}[t]
\includegraphics[width=0.9\linewidth]{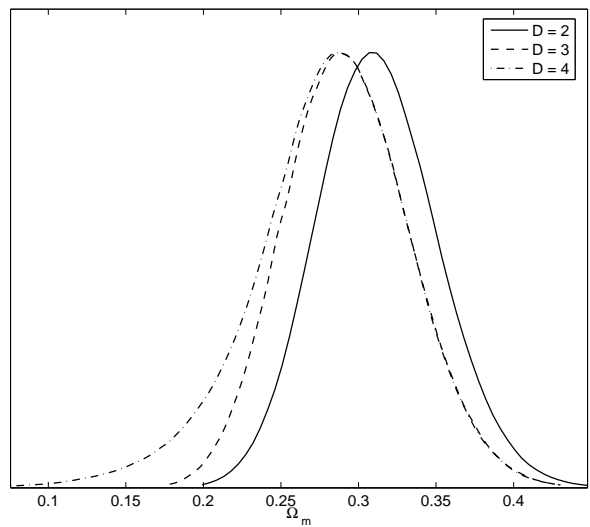}
\caption{Derived marginalised distributions for $\Omega_{\rm m}$.}
\label{fig:omegam_combo}
\end{figure}

Since a non-zero kinetic contribution is preferred by the data, we
also require a higher $V_0$ than in the cosmological constant case. A
simple way to see this should be the case is by considering the
effective quintessence equation of state: with a kinetic contribution
which increases with redshift, the potential term must be larger than
for the cosmological constant case to maintain the same effective
equation of state at high redshift. The corresponding shift and spread
in $\Omega_{\rm m}$ is shown in \reffig{fig:omegam_combo}.

\begin{figure}[t]
\subfigure[\,\,Posterior distribution.]{
\includegraphics[width=0.45\linewidth]{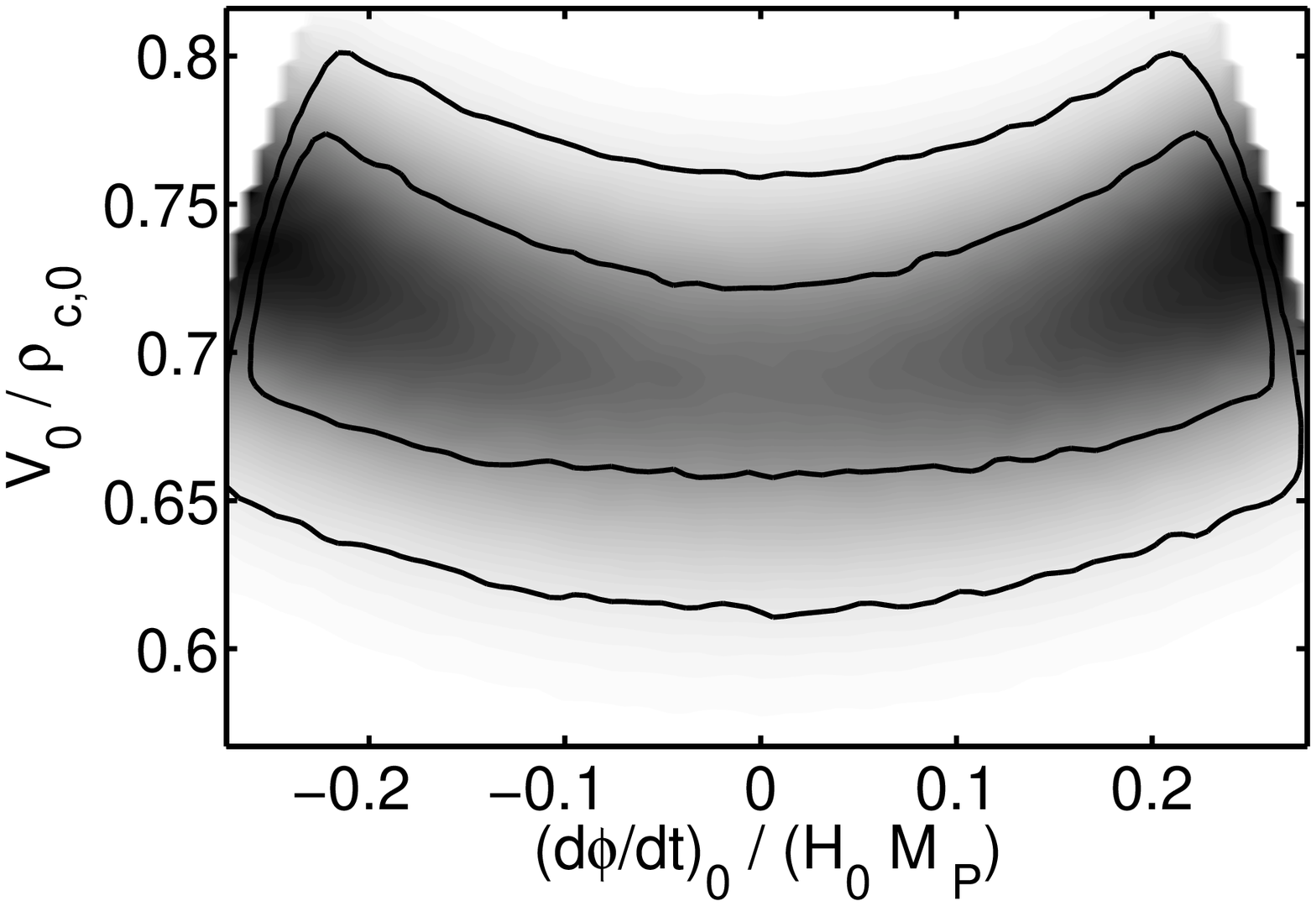}}
\subfigure[\,\,Prior distribution. Shaded region is forbidden.]{
\includegraphics[width=0.45\linewidth]{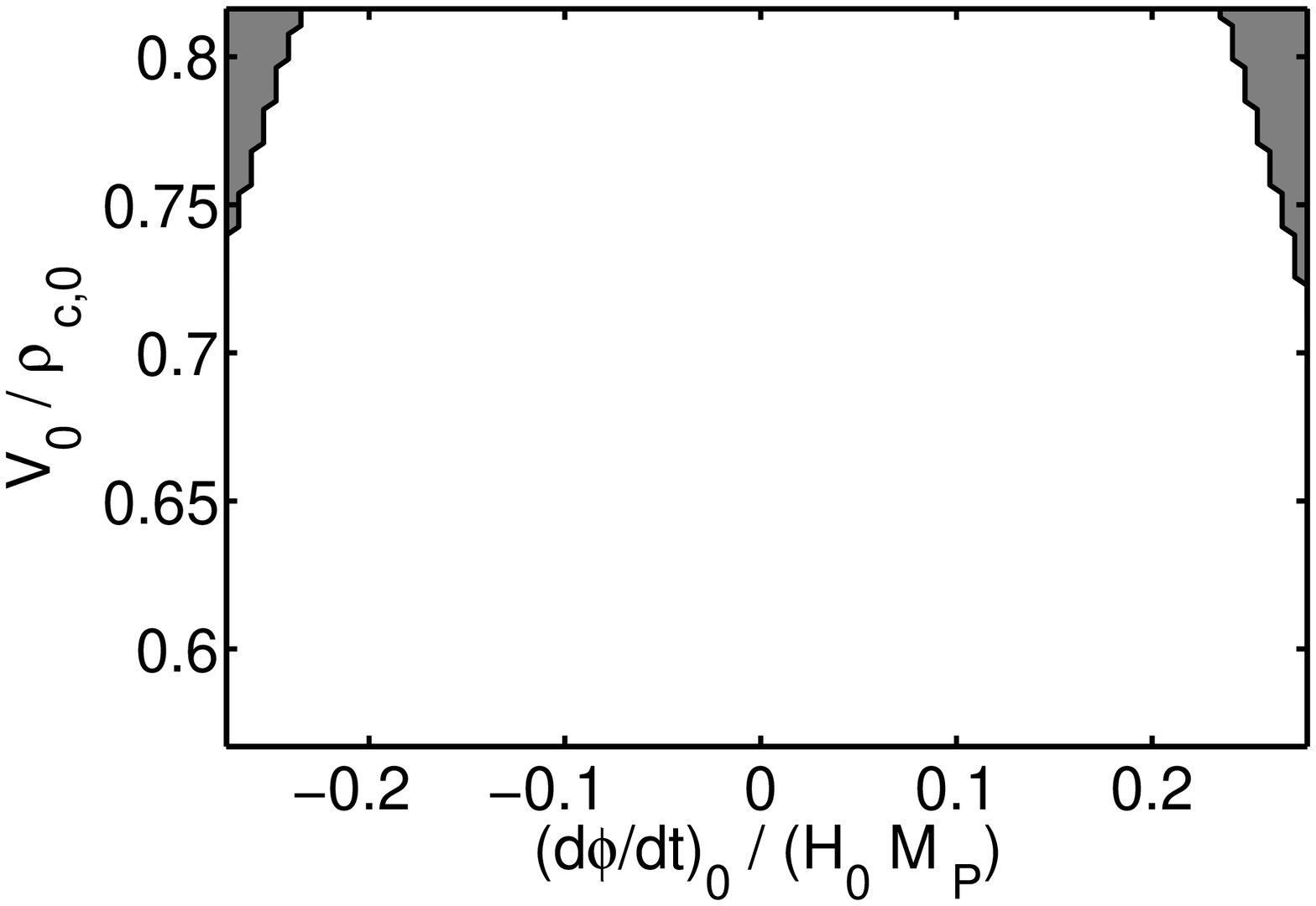}}
\caption{$D=3$. Posterior and prior distributions for $V_0$ and
$\dot{\phi}_0$.}
\label{fig:d3_prior}
\end{figure}

The limits on $\dot{\phi}_0$ (and also the other parameters) are
dependent on our choice of prior on $\Omega_{\rm kin}$, but the above
conclusions remain even in the (unrealistic) case of no prior. Hence,
the choice of prior on $\Omega_{\rm kin}$ can be effectively regarded
as a choice of upper limit on $|\dot{\phi}_0|$.  The cut-off of the
prior on $\Omega_{\rm kin}$ is illustrated in \reffig{fig:d3_prior}.

\subsubsection{Linear potential ($D=4$)}

\begin{figure}[t]
\includegraphics[width=\linewidth]{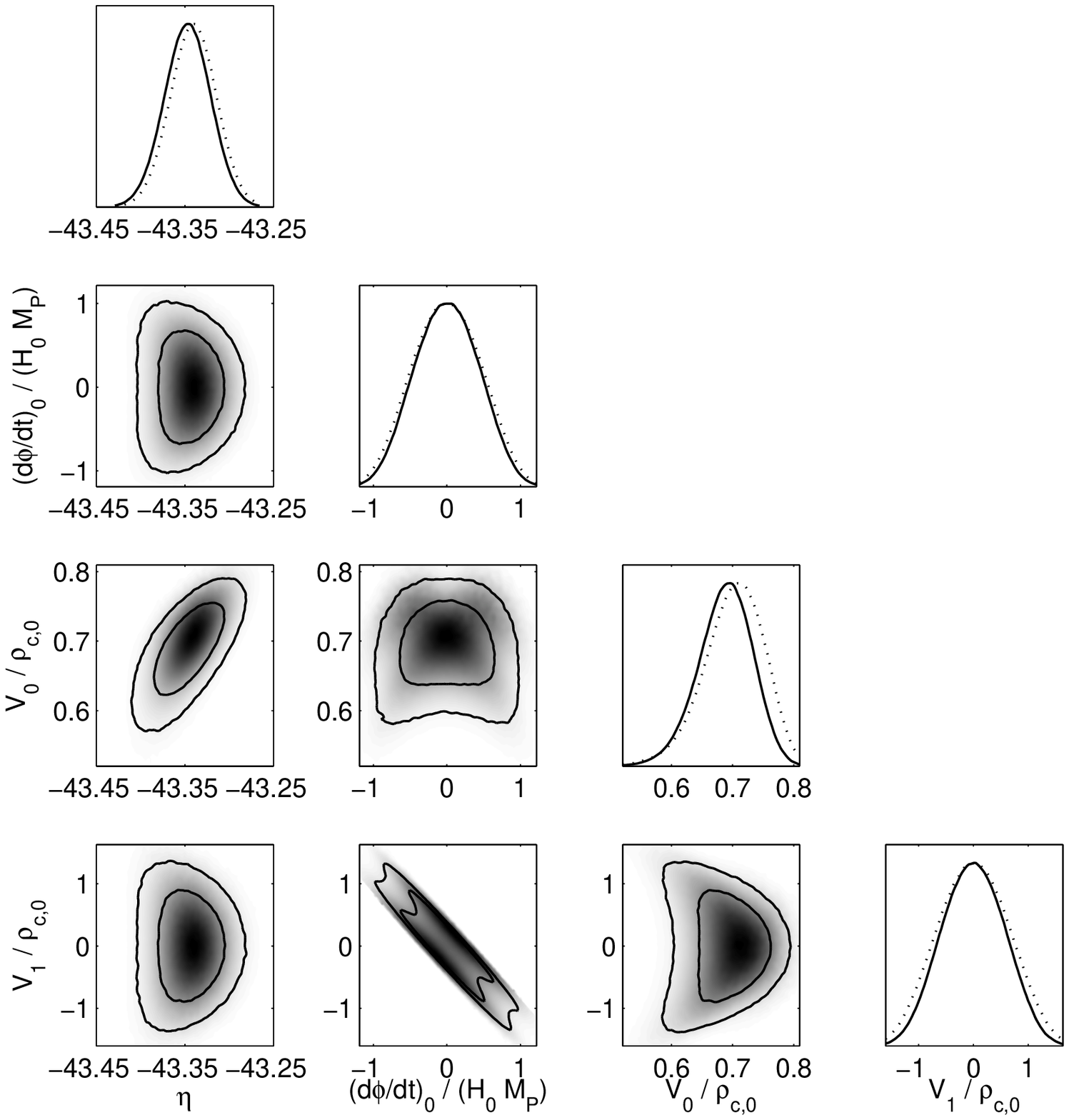}
\caption{As \reffig{fig:d2_tri} for $D=4$.}
\label{fig:d4_tri}
\end{figure}

For a linear potential, the parameter vector under consideration is
\begin{equation}
  \mathbf{\Theta} = (\eta, \dot{\phi}_0, V_0, V_1) \,.
\end{equation}
The likelihood distributions (see \reffig{fig:d4_tri}) show a strong
degeneracy between $\dot{\phi}_0$ and $V_1$, which is the main new
feature compared to $D=3$. This is because a particular value of
$\dot{\phi}$ at some earlier redshift can be attained by adjusting
either $\dot{\phi}_0$ or $V_1$.  Consequently, the best-fit value for
$\dot{\phi}_0$ is less than for $D=3$, but with a non-zero $V_1$.

\begin{figure}[t]
\subfigure[\,\,Posterior distribution.]
  { \includegraphics[width=0.45\linewidth]{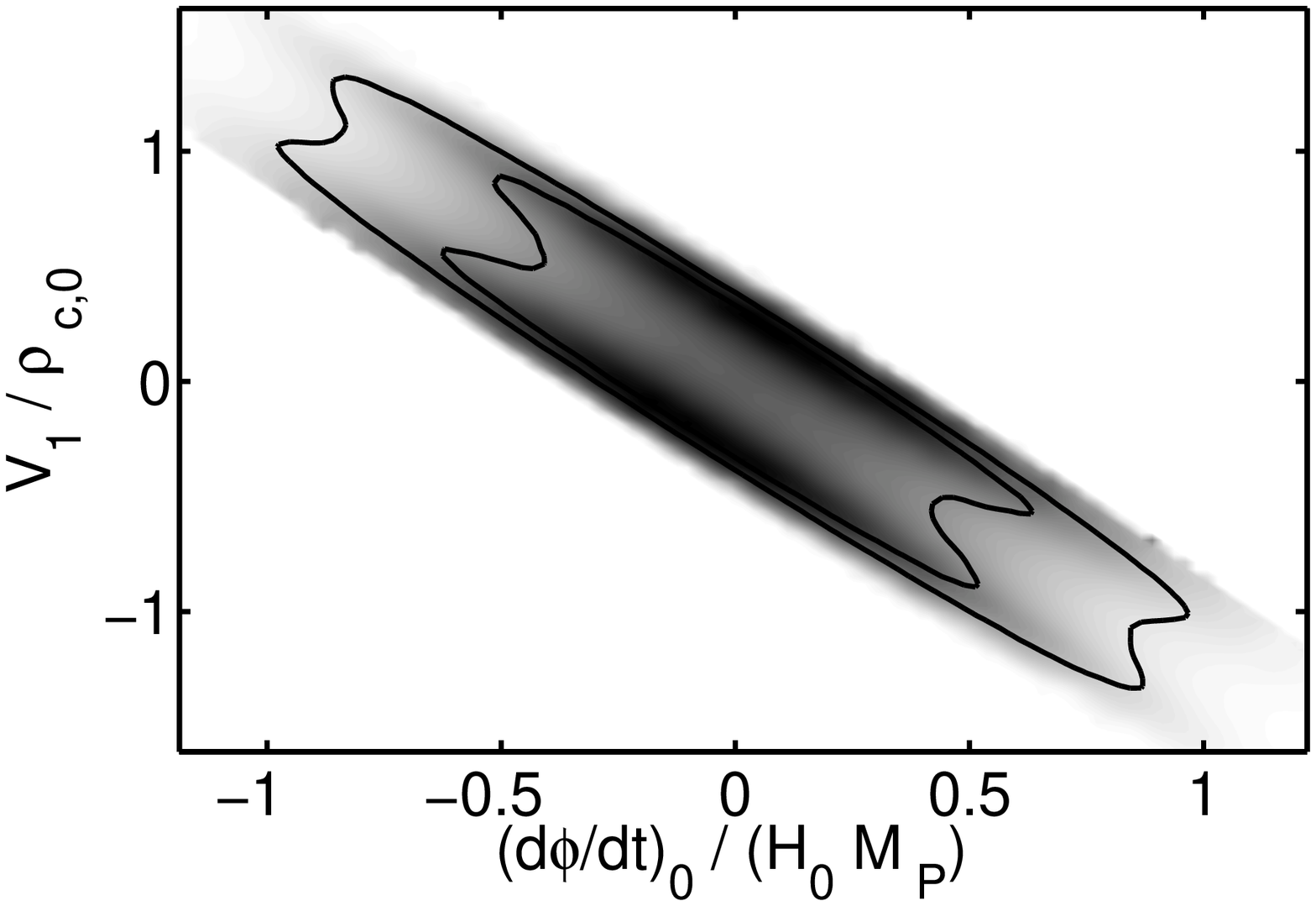}}
  \subfigure[\,\,Prior distribution. Shaded region is forbidden.]{
  \includegraphics[width=0.45\linewidth]{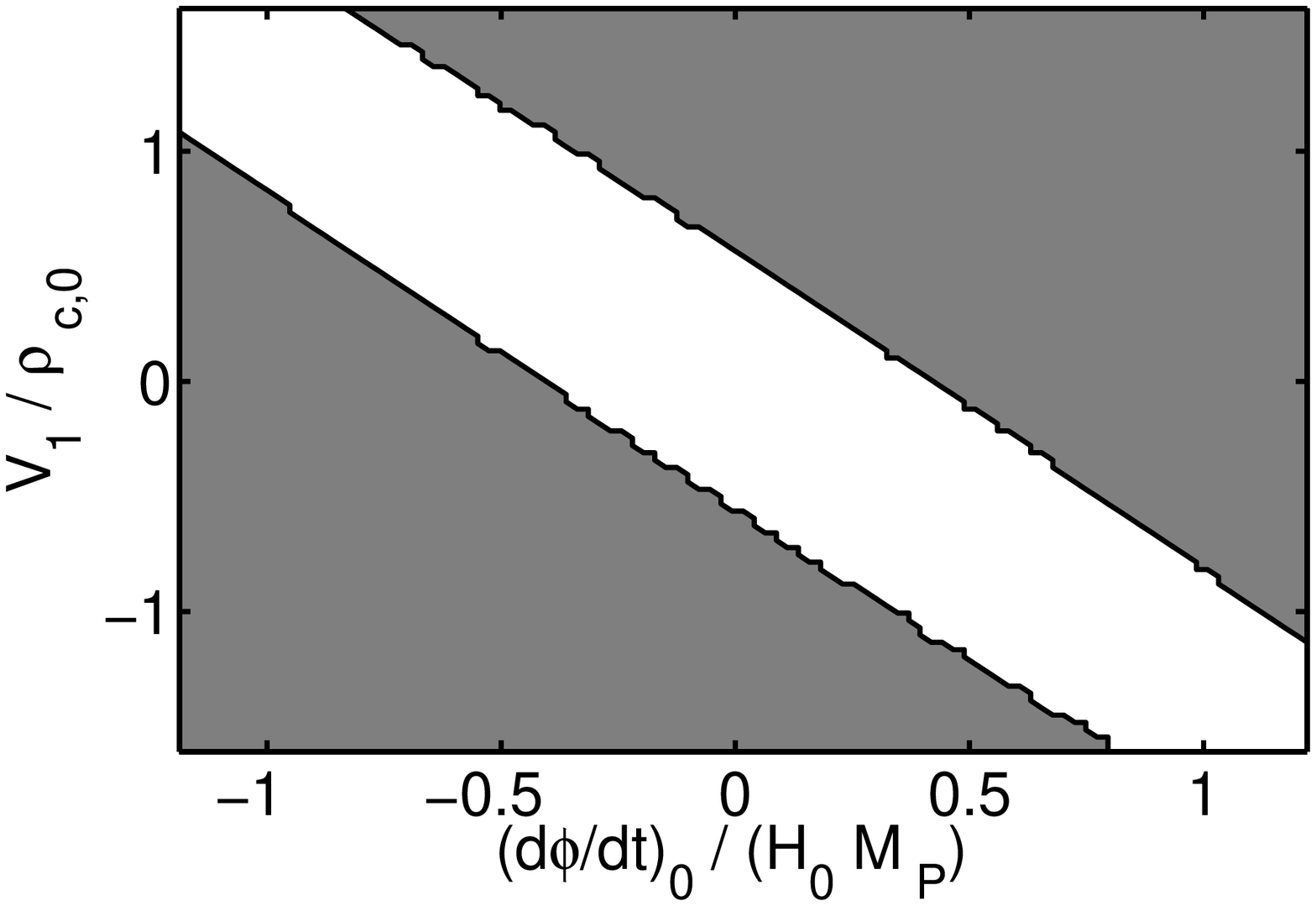}}
\caption{$D=4$. Posterior and prior distributions for $V_1$ and
$\dot{\phi}_0$.}
\label{fig:d4_prior}
\end{figure}

We also note the bimodality in the $\dot{\phi}_0 - V_1$
distribution. This reflects the symmetry under simultaneous change of
sign of $\dot{\phi}_0$ and odd-order expansion coefficients mentioned
in the introduction.  Just as for the case $D=3$, the prior on
$\Omega_{\rm kin}$ cuts off the likelihood distribution in a
high-likelihood region (see \reffig{fig:d4_prior}).

An investigation of the linear potential with a different emphasis can be found 
in Ref.~\cite{Peri}.

\begin{table*}
\begin{threeparttable}
  \begin{tabular}{|l|c|c|c|c|}
    \hline
   & $\mathbf{D=2}$ & $\mathbf{D=3}$ & $\mathbf{D=4}$ & $\mathbf{D=5}$ 
\tnote{c} ~~
\\
    \hline\hline
    $\eta$ & $-43.34^{+0.04}_{-0.04}$ & $-43.34^{+0.04}_{-0.04}$ & 
$-43.34^{+0.05}_{-0.05}$ & $-43.32$ \\
    \hline
     $\dot{\phi}_0 / (H_0 M_{\rm P})~\tnote{a}$ & $-$ & 
\parbox{3.5cm}{~\\$0.25$\\ $|\dot{\phi}_0| / (H_0 M_{\rm P}) 
\lesssim 0.29$\\~} 
& 
\parbox{3.5cm}{$0.15$\\ $|\dot{\phi}_0| / (H_0 M_{\rm P}) \lesssim 1.3$} & 
$-0.07$\\
    \hline
    $V_0 / \rho_{\rm c,0}$ & $0.69^{+0.06}_{-0.06}$ & $0.74^{+0.06}_{-0.10}$ 
& 
$0.73^{+0.07}_{-0.13}$ & $0.96$ \\
    \hline
    $V_1 / \rho_{\rm c,0}$ \tnote{b} & $-$ & $-$ & \parbox{3.5cm}{~\\$0.13$\\ 
$|V_1| / \rho_{\rm c,0} \lesssim 1.8$\\~} & $1.99$ \\
    \hline
    $V_2 / \rho_{\rm c,0}$ & $-$ & $-$ & $-$ & $4.50$ \\
    \hline
    $- 2 \ln {\cal L}_{\rm max}$ & $177.1$ & $176.0$ & $176.0$ & $173.4$ \\
    \hline
    $\mathrm{BIC}_{D}$ & $187.2$ & $191.2$ & $196.2$ & $198.7$ \\
      \hline
    $\mathrm{BIC}_{D} - \mathrm{BIC}_2$ & $0$ & $4.0$ & $9.0$ & $11.5$\\
    \hline
  \end{tabular}
    \begin{tablenotes}
\item[a] The likelihood distribution is an even function of
$\dot{\phi}_0$. No confidence limits can be given for $\dot{\phi}_0$,
as the prior on $\Omega_{\rm kin}$ cuts in the high-likelihood region
and $\dot{\phi}_0 = 0$ is not excluded at the $68\%$ level. The upper
limit on $|\dot{\phi}_0|$ thus corresponds to the maximum allowed
value according to our choice of prior on $\Omega_{\rm kin}$, as
discussed in the text, the quoted number being the best-fit.  
\item[b] The likelihood distribution is
symmetric under simultaneous change of sign of $\dot{\phi}_0$ and
odd-order potential expansion coefficients. No confidence limits can
be given for $\dot{\phi}_0$ or $V_1$ for the same reason as above.
\item[c] Because of the difficulty in obtaining a
convergent/well-mixed sampling, as discussed in the text, we choose
not to give confidence limits for $D=5$.  \end{tablenotes}
\end{threeparttable}
\caption{Best-fit model parameters and BIC values. Note that these are
the best-fit parameter values and confidence limits derived from the
full $D$--dimensional likelihood distribution, \emph{not} the
marginalised distributions.}  
\label{tab:bf}
\end{table*}

\subsubsection{Quadratic potential ($D=5$)}

The quadratic potential model has the parameter vector
\begin{equation}
  \mathbf{\Theta} = (\eta, \dot{\phi}_0, V_0, V_1, V_2) \,.
\end{equation}
In this case we find that the third potential parameter, $V_2$, is
unconstrained by the data, characterised by a large and oscillating
Gelman--Rubin ratio (around 1.3--1.9) for that parameter. Because of
that we do not show the likelihoods, since the marginalised
distributions will not have the correct weights.

To explore this situation further, we ran additional chains using
$\tilde{V}_2 \equiv \arctan(V_2)$ as our parameter instead of $V_2$
(this corresponds to a change in prior on $V_2$, since $\de
\tilde{V}_2/\de V_2$ is a function of $V_2$).  This choice was
motivated by the expectation that a cosmological constant, which is
achieved in either of the limits $V_2=0$ or $V_2=\infty$, is very
likely to be a good fit, and hence allows us to explore the infinite
range in $V_2$ that might be needed.  With this choice we get
convergent and low Gelman--Rubin ratios, and $\arctan(V_2)$
essentially unconstrained.  At $V_2 \sim \rho_{{\rm c},0}$ we find a
small peak in likelihood, but because the distribution remains high
and nearly flat outside this peak it is not possible to constrain the
parameter without further data.

\subsection{Model comparison}

We compare the different models using the BIC, which uses the maximum
likelihood achievable by each model. The parameters, likelihoods and
BIC values are given in \reftab{tab:bf}.

\subsubsection{Cosmological constant ($D=2$)}

The cosmological constant forms the base model for our model
comparison, and as is well known provides a good fit to the
data. Indeed, the BIC ranks it as preferred to our other models.

\begin{figure*}
    \begin{tabular}{|ccccc|}
      \hline
       & $\mathbf{D=2}$ & $\mathbf{D=3}$ & $\mathbf{D=4}$ & $\mathbf{D=5}$ \\
\includegraphics[width=0.02\linewidth,bb=28 40 91 414]{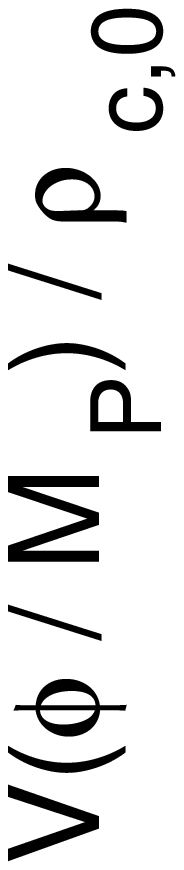} &
\includegraphics[width=0.19\linewidth]{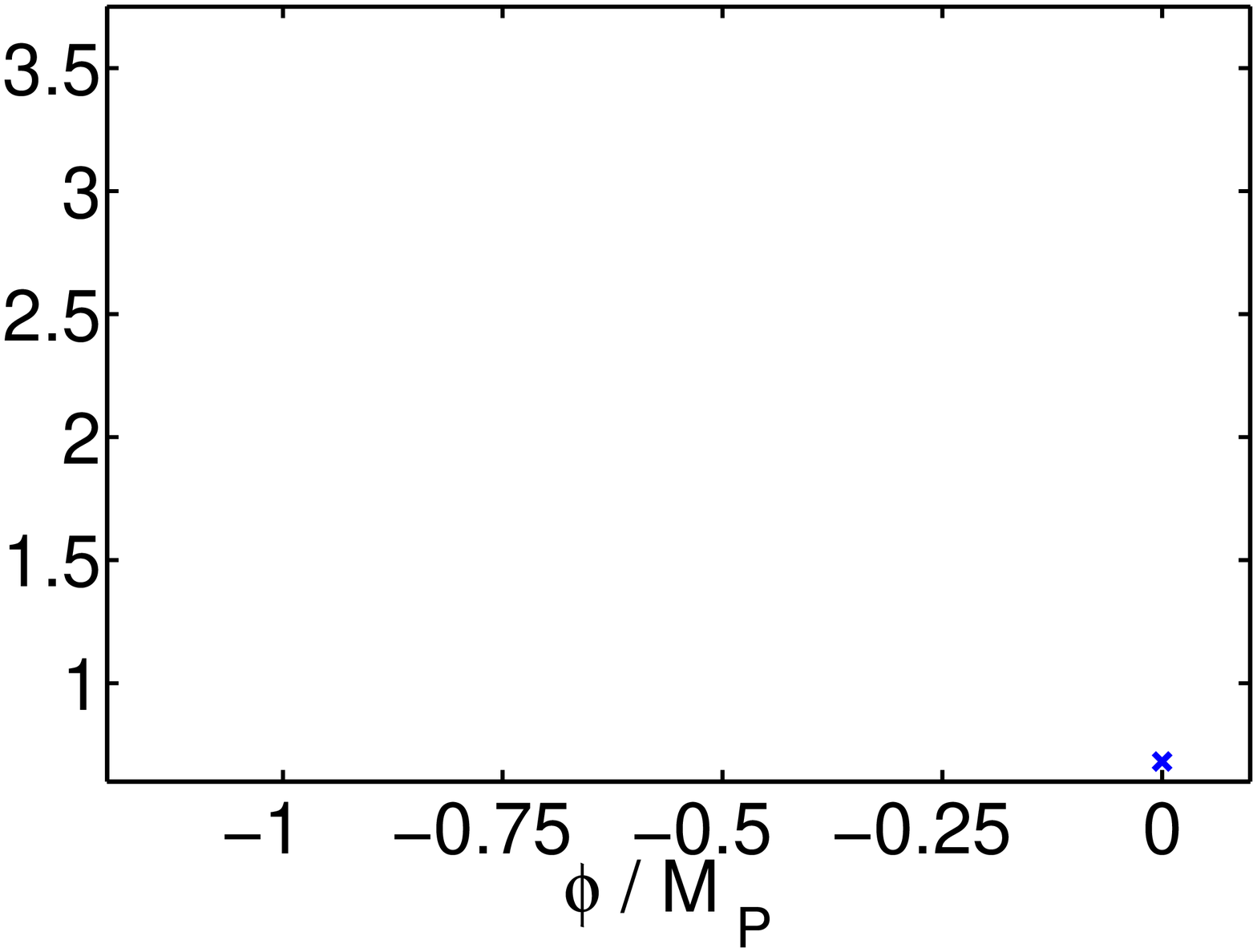} &
\includegraphics[width=0.19\linewidth]{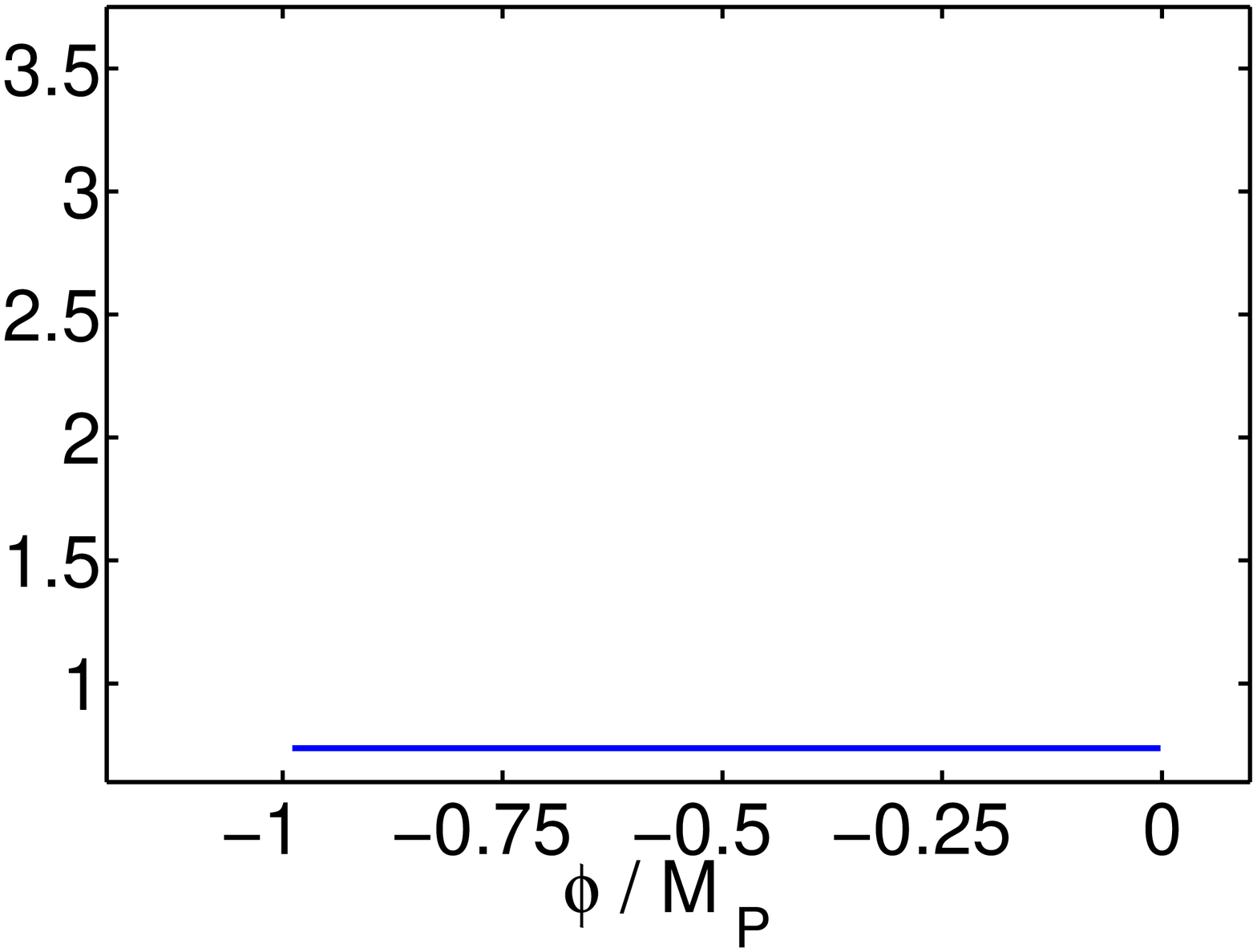} &
\includegraphics[width=0.19\linewidth]{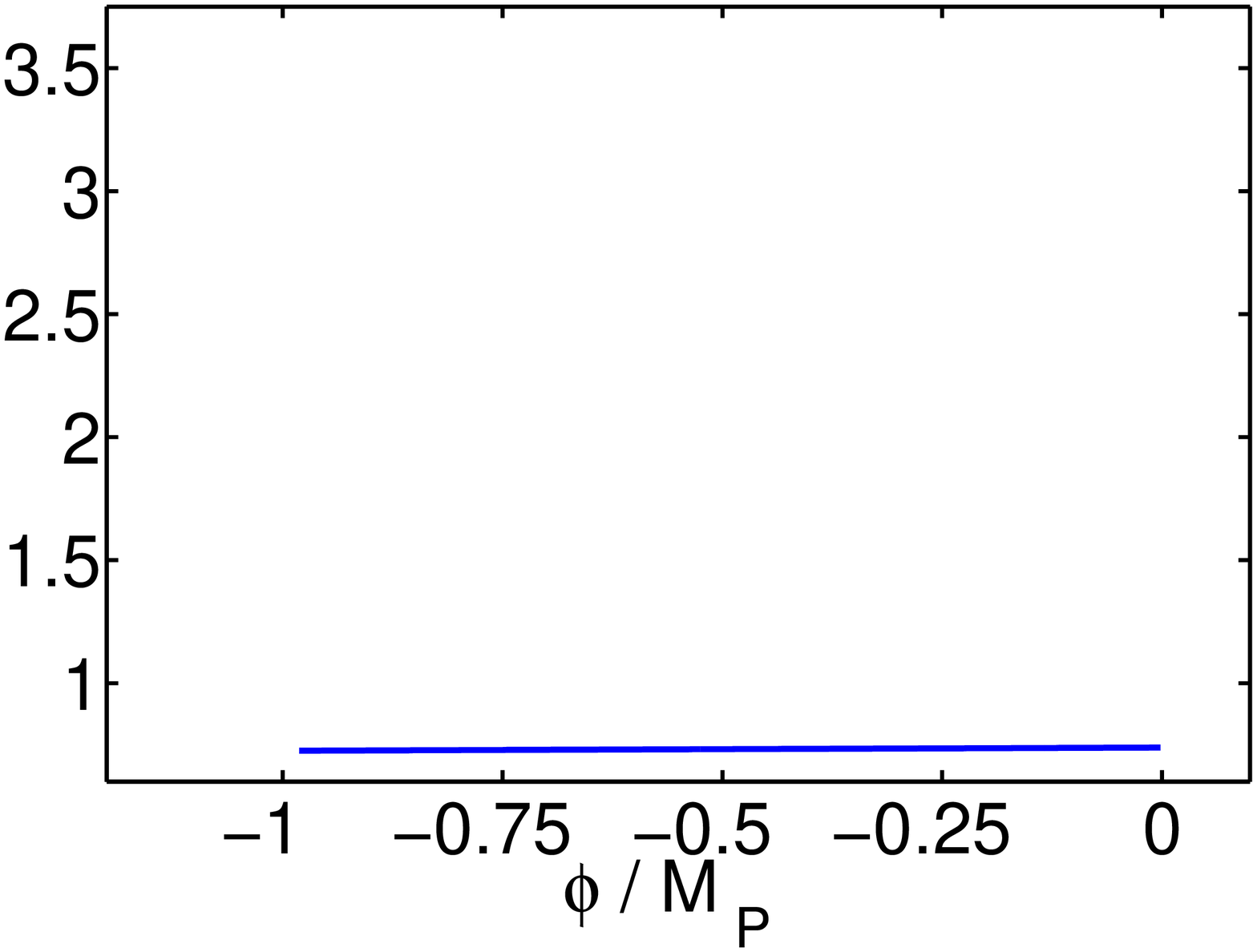} &
\includegraphics[width=0.19\linewidth]{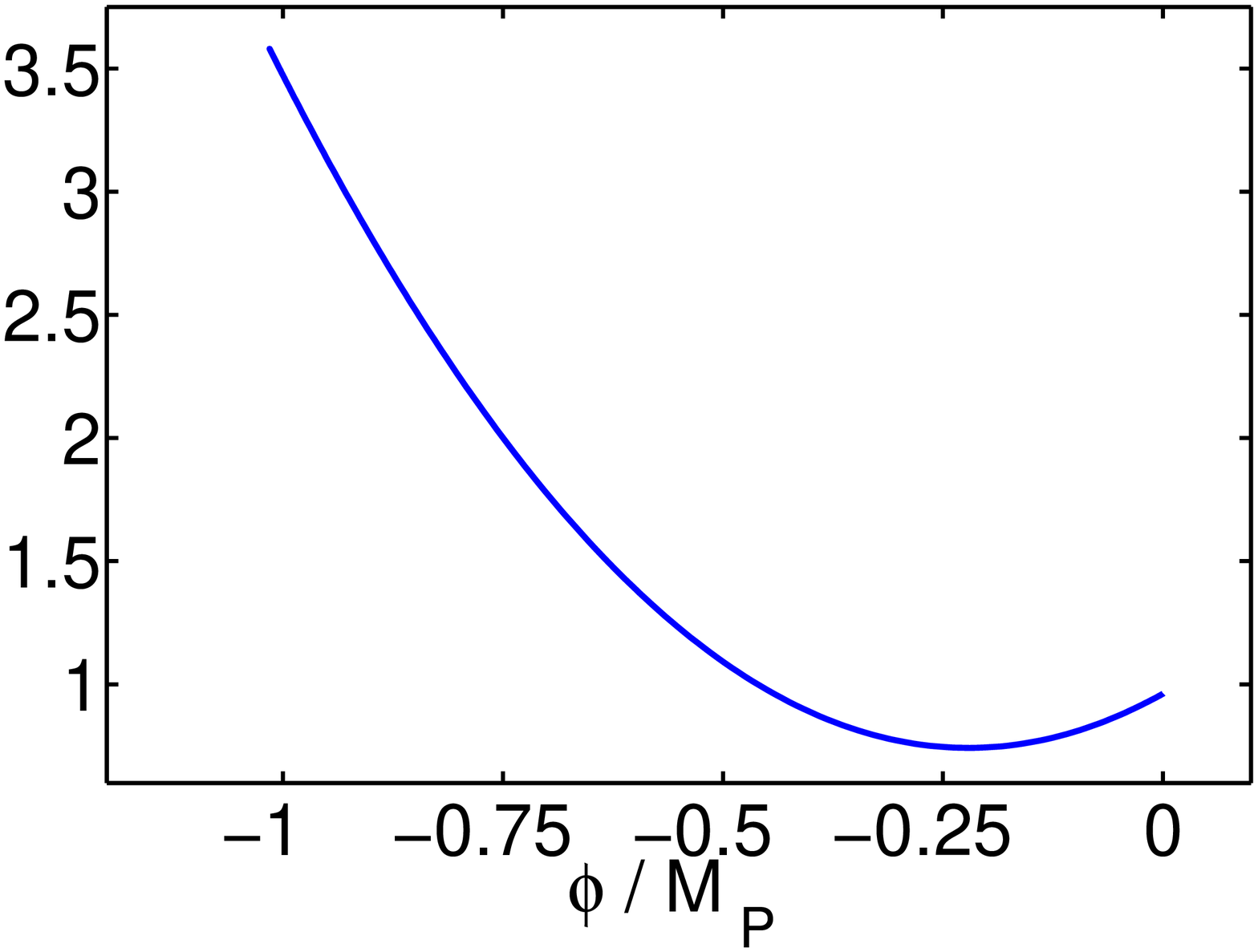} 
\\
\includegraphics[width=0.02\linewidth,bb=3 375 68 1753]{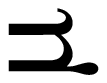} &
\includegraphics[width=0.2\linewidth]{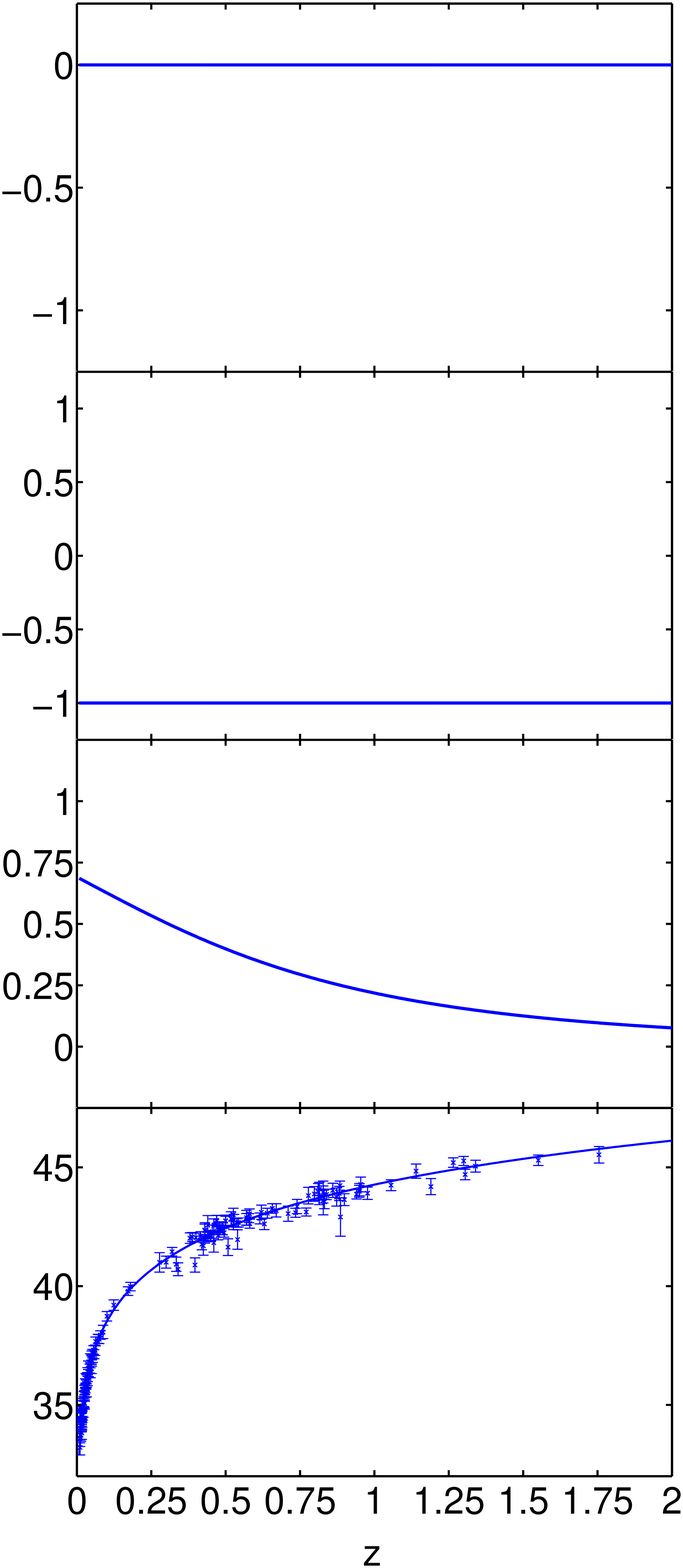} &
\includegraphics[width=0.2\linewidth]{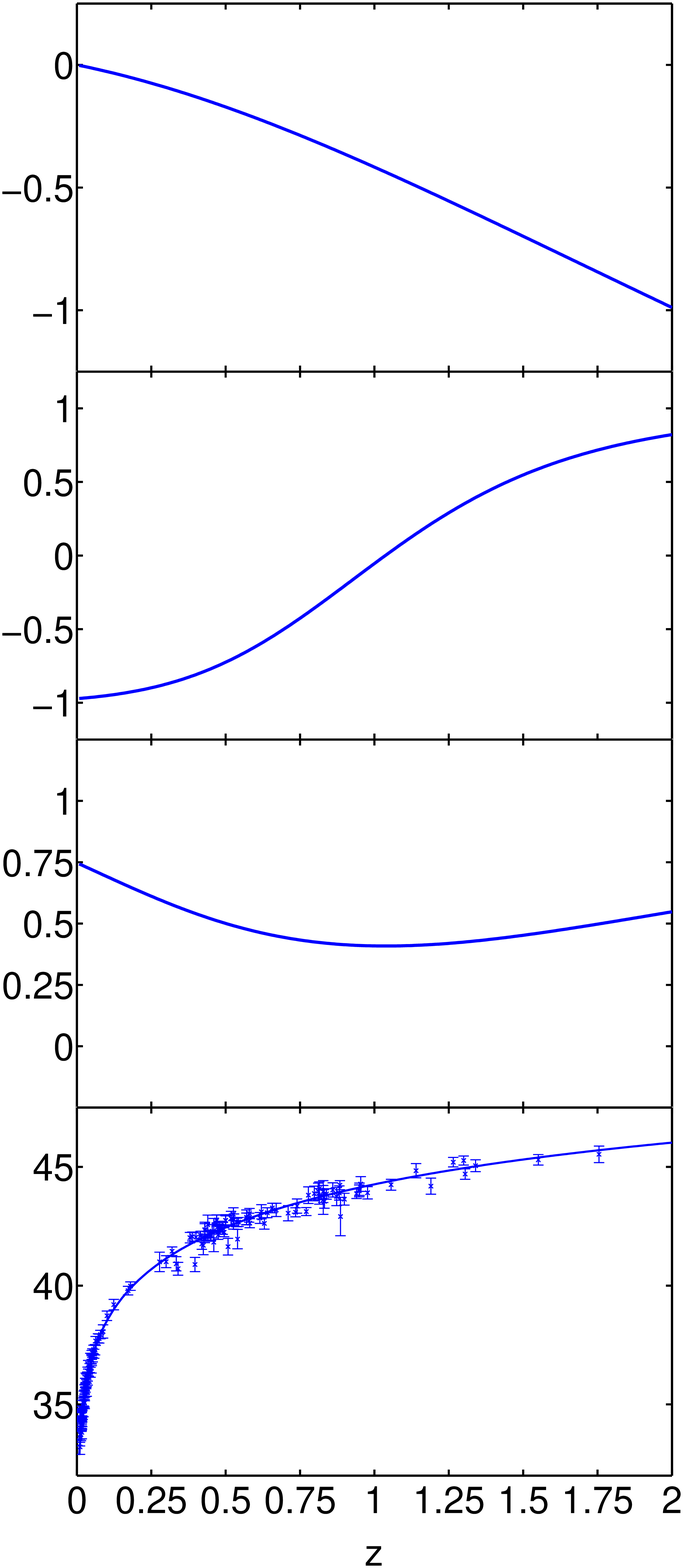} &
\includegraphics[width=0.2\linewidth]{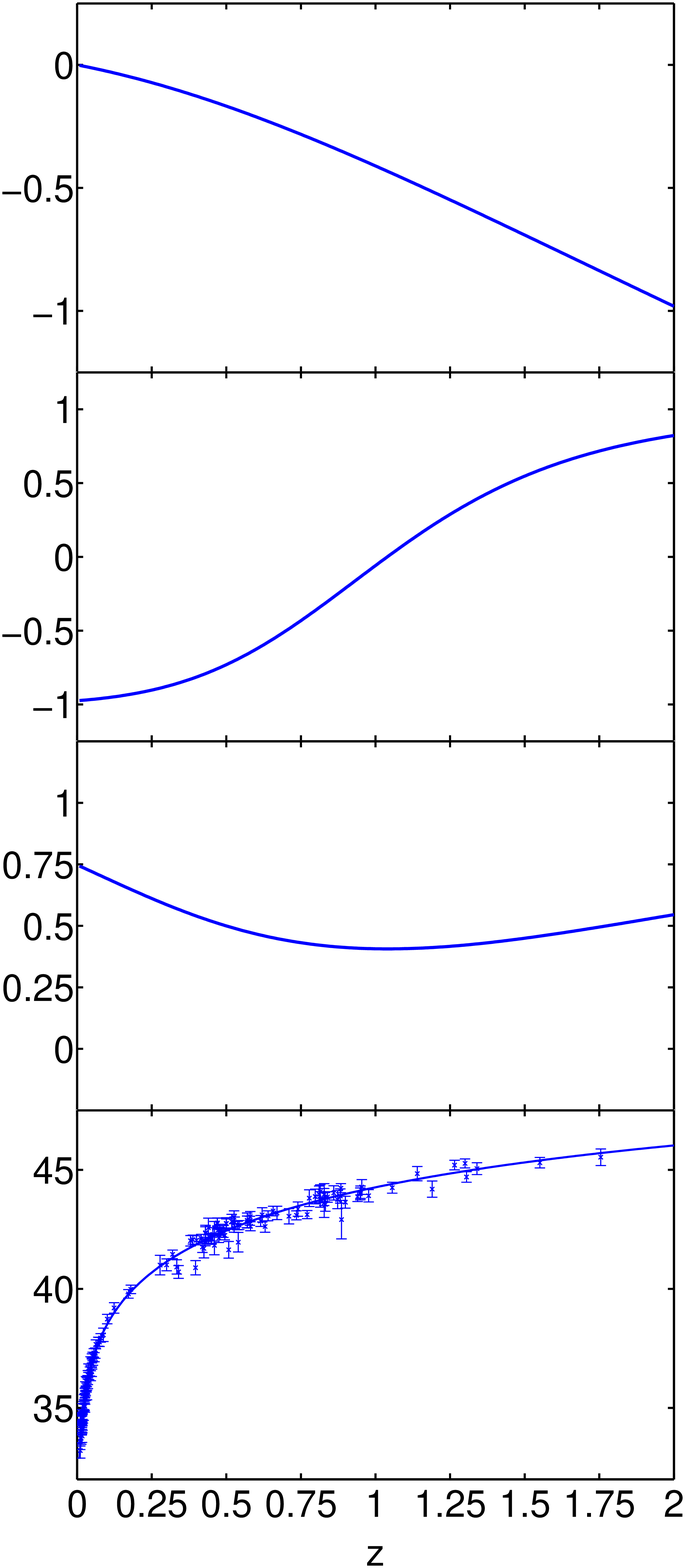} &
\includegraphics[width=0.2\linewidth]{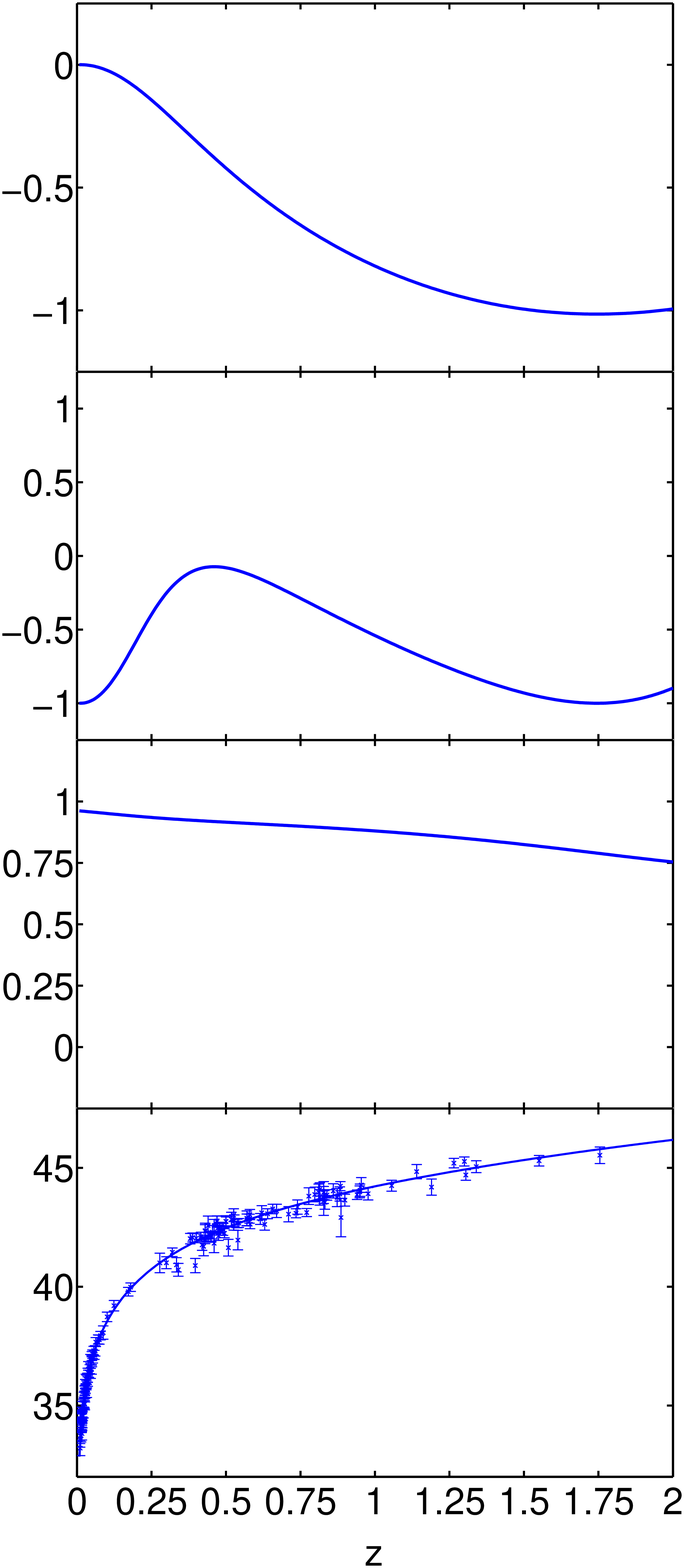} 
\\
\hline
    \end{tabular}
\caption{Dynamical evolution in the best-fit cosmologies. The graphs of the 
potential show the range of $\phi$ out to $z=2$.}
\label{fig:bf}
\end{figure*}

\subsubsection{Constant potential with kinetic energy ($D=3$)}

As mentioned in the context of parameter fitting, $\dot{\phi}_0 = 0$
is not excluded at a statistically-significant level. Including this
parameter does allow a somewhat better fit to the data, but the BIC
penalises its extra parameter leaving the pure cosmological constant
model as the preferred description of present data.

The best-fit $D=3$ cosmology, shown in \reffig{fig:bf}, exhibits very
strong evolution in $w_{\phi}$ from kinetic to potential domination
over the redshift range of available data. This is not entirely
surprising: even a tiny kinetic contribution at present will
correspond to a much higher kinetic component at earlier times, as
only Hubble friction will work to decrease $\dot{\phi}$ in this model.

\subsubsection{Linear potential ($D=4$)}

As clearly seen, $\{\dot{\phi}_0=0,V_1=0\}$ is well within the
preferred region, and so the cosmological constant model is embedded
within the allowed parameter space of our extended model. Accordingly,
the model comparison of \reftab{tab:bf} prefers the pure cosmological
constant model, with the BIC difference arguing quite strongly against
the inclusion of the two extra parameters.

The best-fit cosmology (\reffig{fig:bf}) is practically
indistinguishable from the best-fit for $D=3$. This arises from the
strong degeneracy between $\dot{\phi}_0$ and $V_1$; it turns out that
present data are not discriminating in the orthogonal direction.

Within the context of model comparison, a curious point to note is
that a field rolling on a linear potential is quite strongly
disfavoured as compared to a field rolling on a constant potential
(\reftab{tab:bf}). This is because the inclusion of a potential slope
hardly improves the best-fit at all, while costing an extra
parameter. This may seem quite artificial, but is the conclusion of
our phenomenological approach.  One should note however that the BIC
comparison addresses only how well the different models fit the data
at hand; when interpretting as a model probability one should bear in
mind that this conclusion could be overturned if one felt that the
prior model probabilities were quite different.

\subsubsection{Quadratic potential ($D=5$)}

As \reftab{tab:bf} shows, the quadratic model is strongly disfavoured
by the BIC. Note that although we have not necessarily obtained a
convergent distribution, we can assess the model as the BIC only
depends on the \emph{maximum} likelihood value. However, with such a
broad distribution the BIC is expected to be a poor approximation to
the Bayesian evidence.

A significant feature of the quadratic potential model is that the
best-fit has $\Omega_{\rm m} \approx 0.05$, see \reffig{fig:bf}. This
is of course in stark contradiction with many other datasets. The
evolution of $w_{\phi}$ is quite different from that for $D=3$ and
$D=4$, with the field starting high up on the potential, rolling past
the minimum and reaching the turning point by the present
time. However the preferred parameter region includes models with much
more reasonable $\Omega_{{\rm m}}$.

\subsection{Choice of prior on $\Omega_{\rm kin}$}

\label{s:kin}

As mentioned above, we limit the kinetic contribution at $z \ge
1$. This is necessary because the SNIa data alone favourthe kinetic
energy to dominate at $z = 1$, and by inference to be completely
dominant at higher redshifts. This is in contradiction to almost any
other cosmological dataset (for instance, the mere existence of
high-redshift galaxies, and of the cosmic microwave background), and
so external priors are necessary to keep us in the physical regime.

\begin{figure}[t]
\subfigure[\,\,$D=3$. $\Omega_{\rm kin} < 0.25 \,\,\forall\, z \ge 1$.]{
  \includegraphics[width=0.45\linewidth]{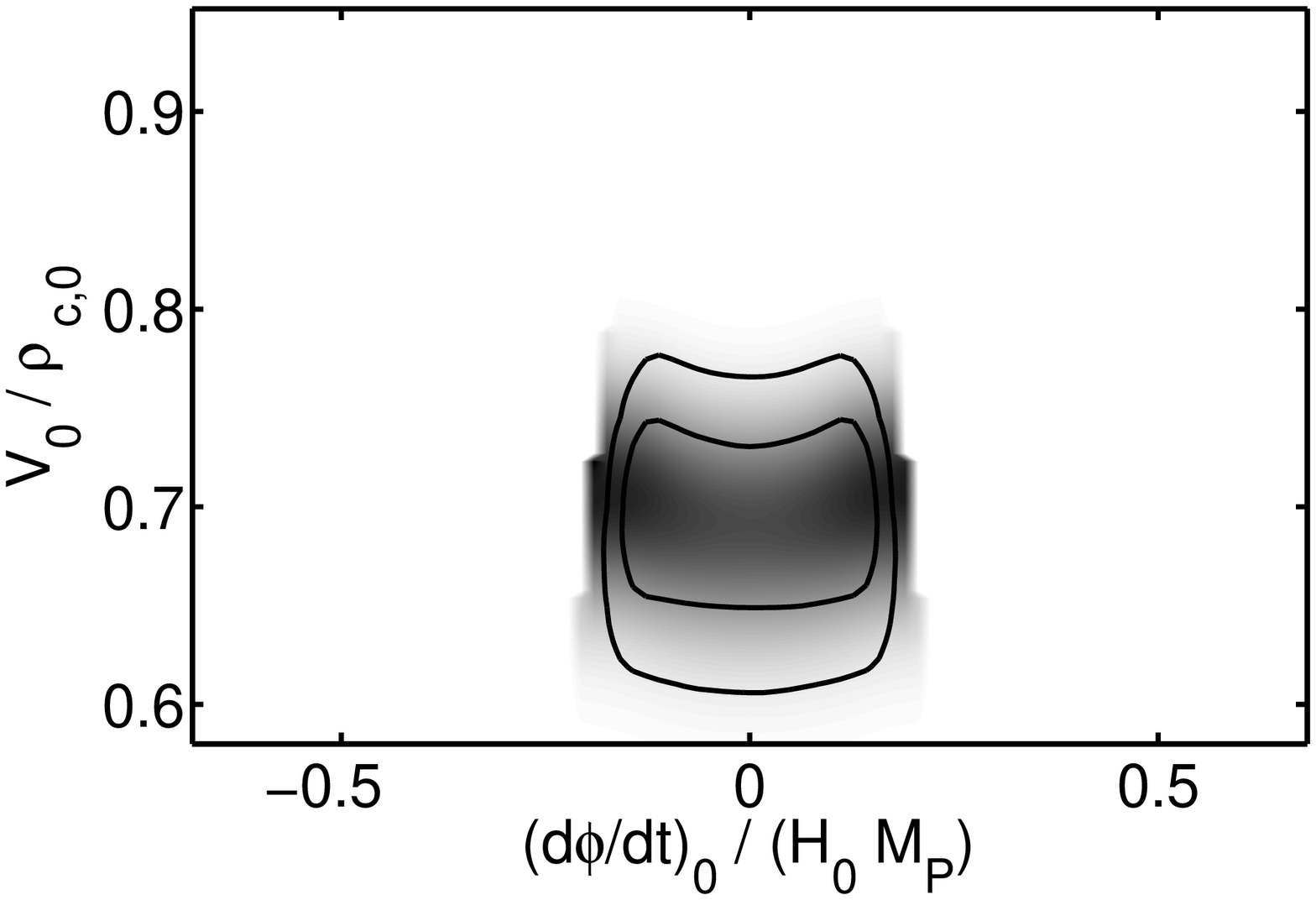}}
\subfigure[\,\,$D=4$. $\Omega_{\rm kin} < 0.25 \,\,\forall\, z \ge 1$.]{
\includegraphics[width=0.45\linewidth]{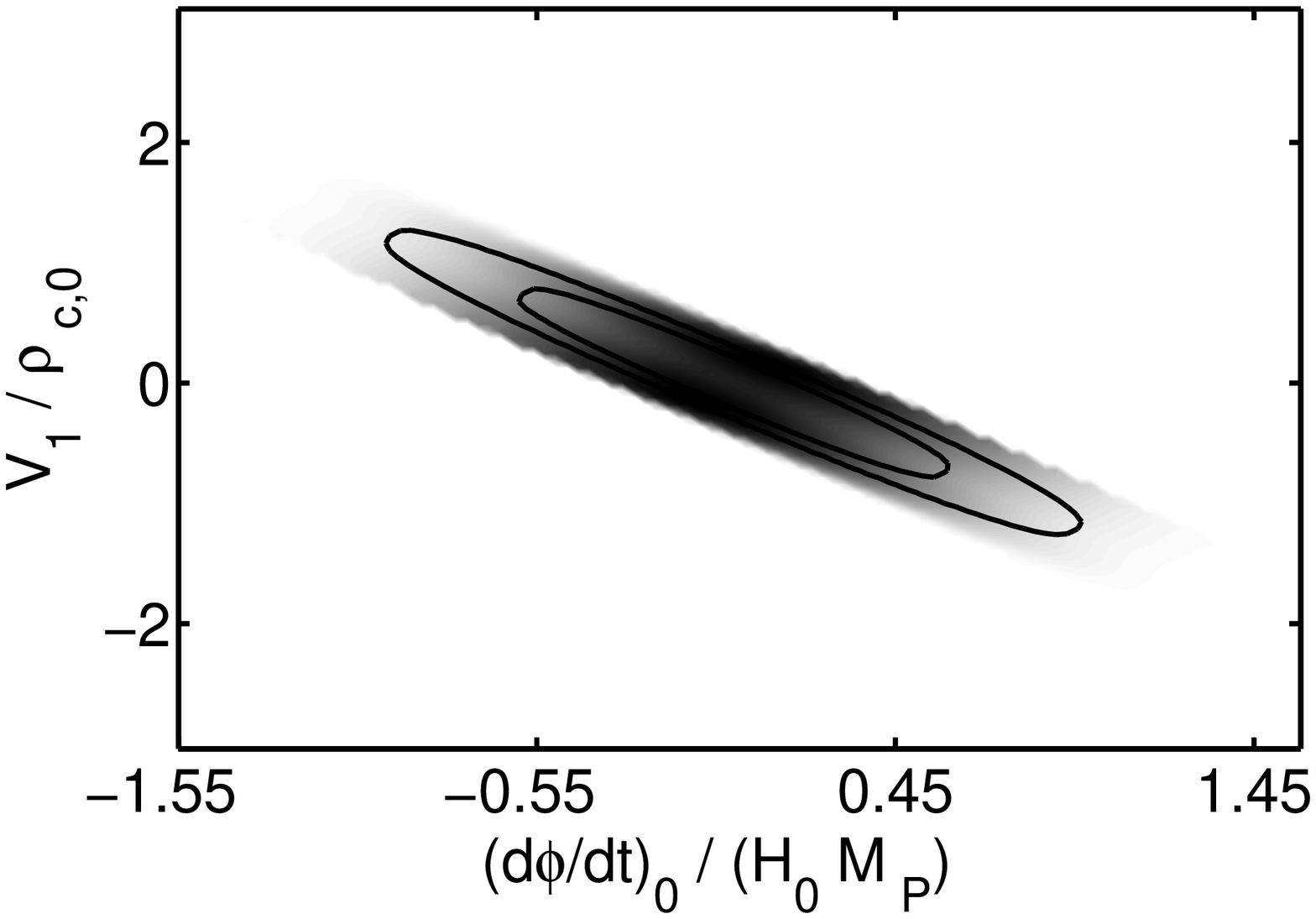}}
\\
\subfigure[\,\,$D=3$. $\Omega_{\rm kin} < 0.5 \,\,\forall\, z \ge 1$.]{
  \includegraphics[width=0.45\linewidth]{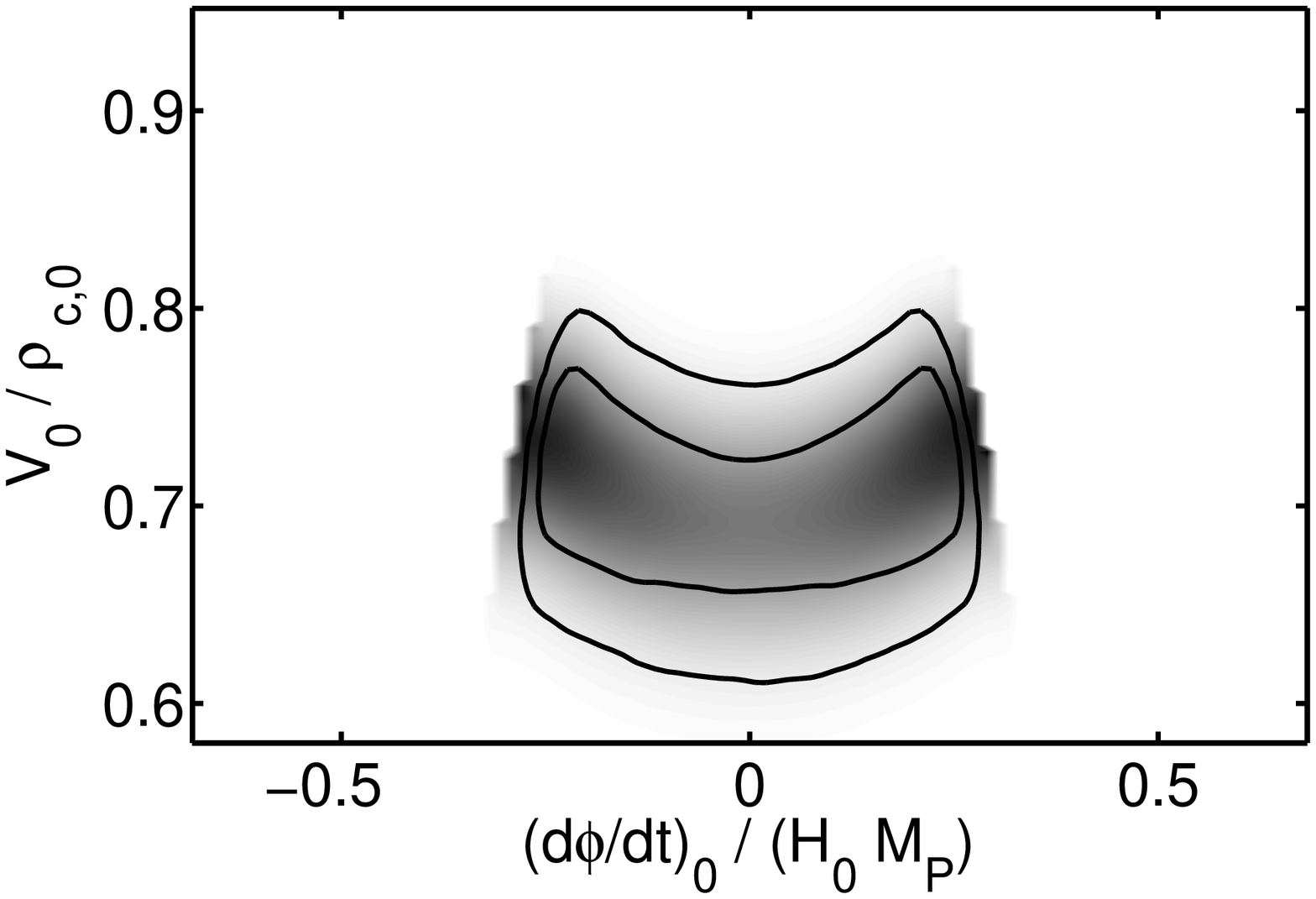}}
\subfigure[\,\,$D=4$. $\Omega_{\rm kin} < 0.5 \,\,\forall\, z \ge 1$.]{
\includegraphics[width=0.45\linewidth]{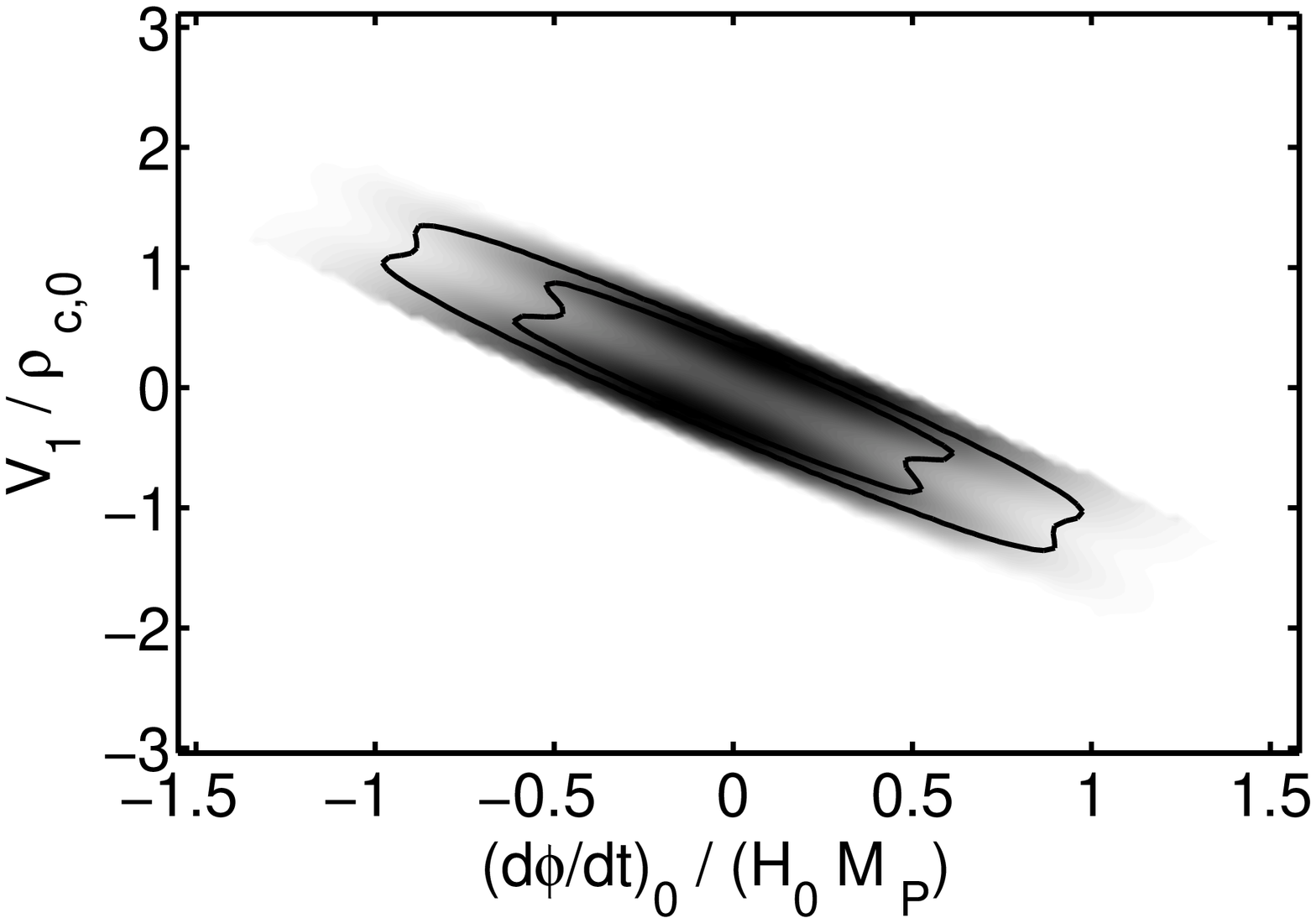}}
\\
\subfigure[\,\,$D=3$. No limit on $\Omega_{\rm kin}$.]{
  \includegraphics[width=0.45\linewidth]{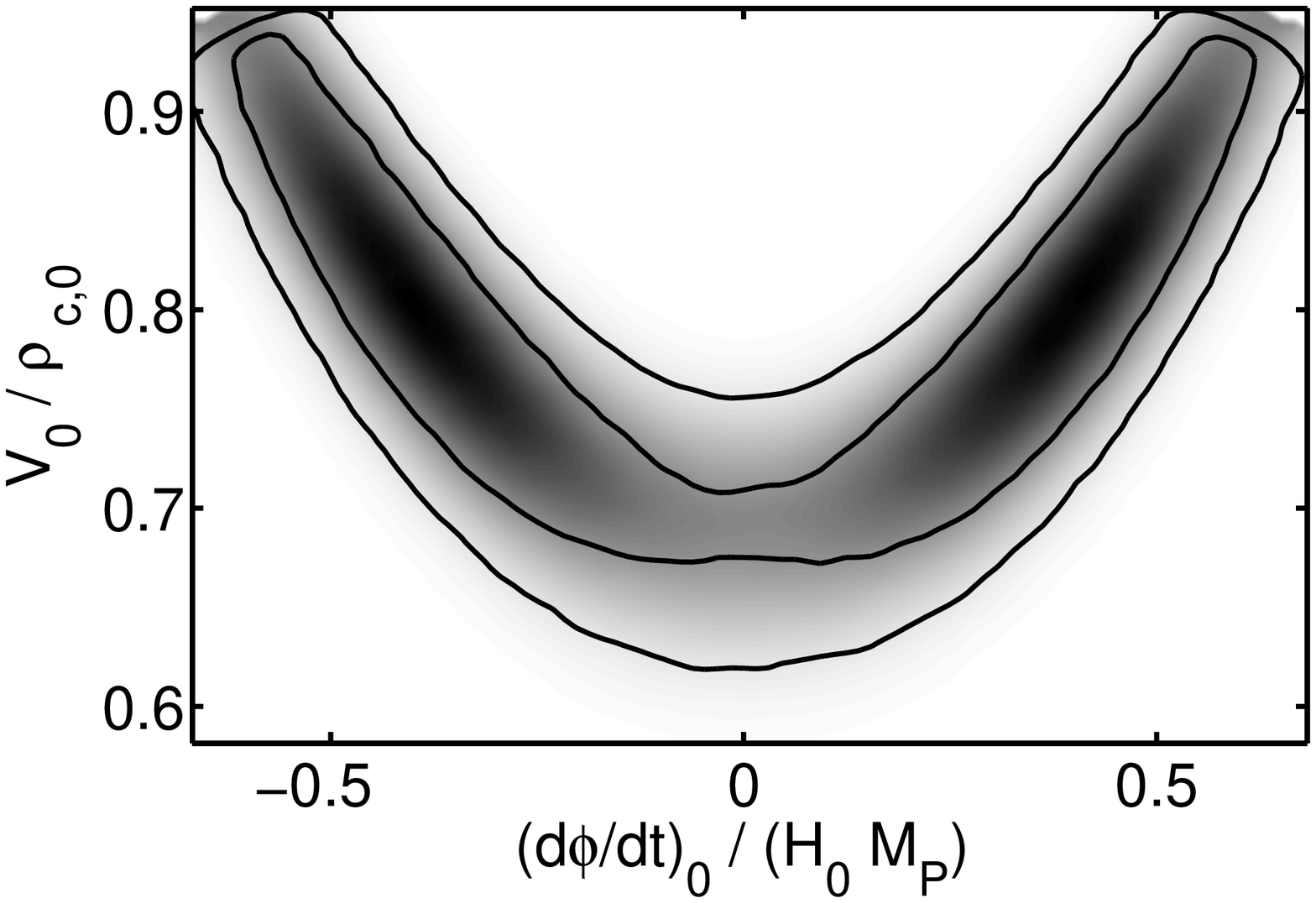}}
\subfigure[\,\,$D=4$. No limit on $\Omega_{\rm kin}$.]{
\includegraphics[width=0.45\linewidth]{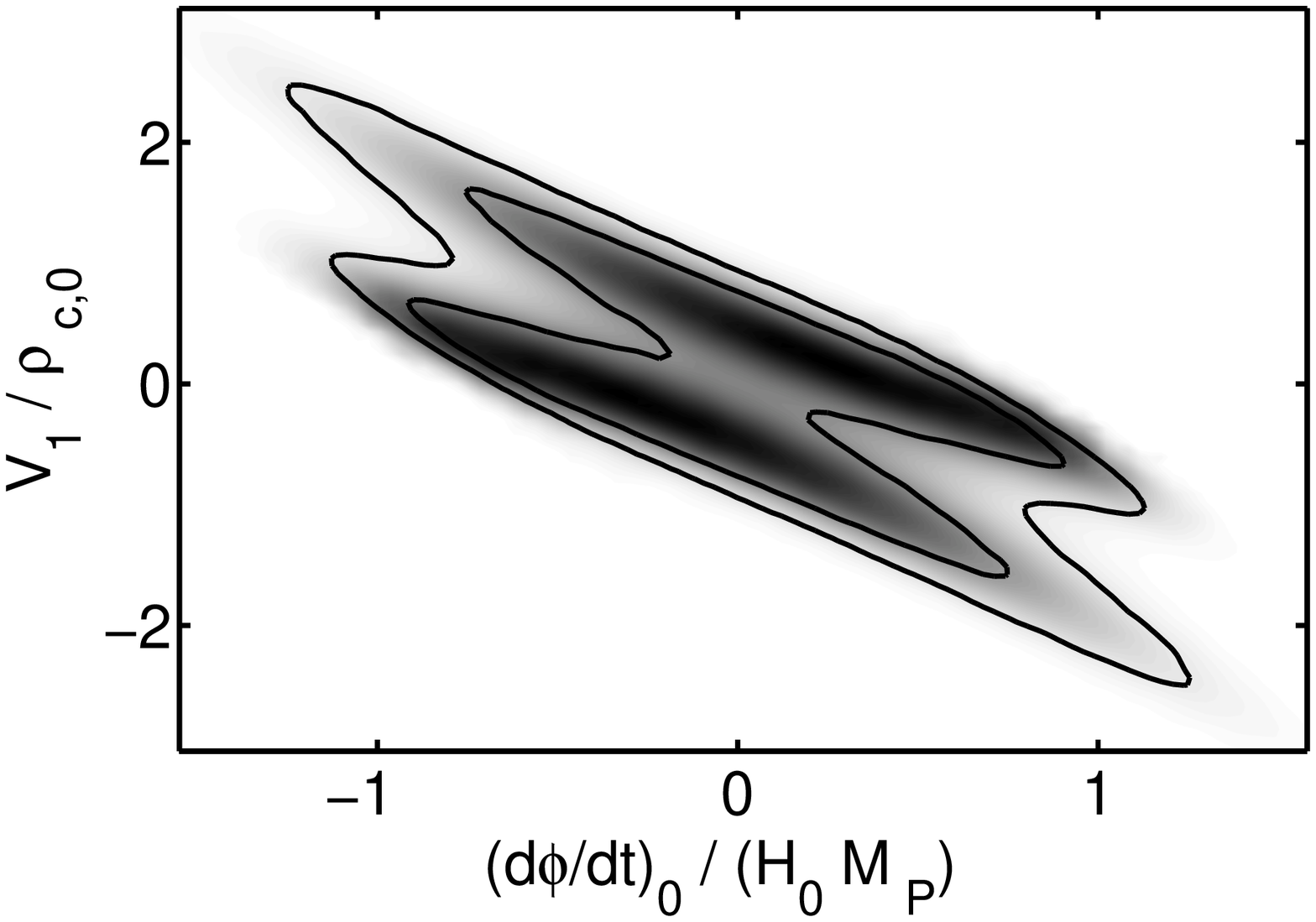}}
\caption{Marginalised posteriors for $D=3$ and $D=4$ depending on
the choice of prior on $\Omega_{\rm kin}$.}
\label{fig:prill}
\end{figure}

The precise choice of this upper limit on $\Omega_{\rm kin}$ is
somewhat arbitrary, and does have a non-negligible impact on the
posterior distribution.  In terms of marginalised distributions, the
distributions involving $\dot{\phi}_0$ show the strongest dependence,
see \reffig{fig:prill}.  Looking at parameter estimation, as mentioned
above the main impact is broader confidence limits.  For our model
selection analysis, conclusions remain unchanged as the cosmological
constant model is still the preferred model even without a prior on
$\Omega_{\rm kin}$ for all cases.

\section{Conclusions}

We have described and implemented a reconstruction scheme for
quintessence potentials from data, using an MCMC likelihood approach,
which we applied to SNIa data. Additionally, we describe the
application of model selection using the Bayesian Information
Criterion, and point out the generality of any positive evidence found
for dynamical dark energy in this approach.

As might be expected, the data provides positive evidence in favour of
a cosmological constant in our setup, based on model selection by the
BIC. A similar conclusion was previously reached by Saini {\it et
al.}~\cite{SWB} and by Bassett {\it et al.}~\cite{BCK} amongst a set
of models parametrised by different equation of state evolution. Some
of our distributions do however exhibit broad non-Gaussian regions,
which merits a more detailed model selection investigation using the
full Bayesian evidence (since the BIC is only a reasonable
approximation for sharply-peaked unimodal distributions).

The low-dimensional models are quite well constrained by current
data. However, we find that if one allows power series of order two or
higher, the parameters (including the linear one) become unconstrained
by current SNIa data given our very loose priors.  This agrees with
what is suggested by the analysis by Maor and Brustein \cite{Maor:2002rd} in the 
context of distinguishing potential
classes.

We plan to extend this work in several directions. As mentioned above,
it would be interesting to extend the model selection to use the full
Bayesian evidence, though it will then be essential to consider the
issue of prior parameter ranges, which the BIC
sidesteps. Generalisation to include further cosmological datasets is
desirable, especially CMB anisotropies, though that will require the
potential expansion to be valid over a much wider range of
redshifts. Finally, it would be interesting to explore whether in the
context of tracking models it might be possible to eliminate the
parameter $\dot{\phi}_0$, which ought to be determined via an early
tracking regime.


\begin{acknowledgments}

M.S.~was supported at Sussex by a Marie Curie Fellowship of the
European Community programme HUMAN POTENTIAL under contract
HPMT-CT-2000-00096.  A.R.L.~and D.P.~were supported by PPARC. We thank
Mar\'{\i}a Beltr\'an, Pier Stefano Corasaniti, Irit Maor, Claudia
Quercellini and Erandy Ramirez for helpful discussions, and Jon
Urrestilla for computer assistance. We additionally acknowledge use of
the UK National Cosmology Supercomputer funded by Silicon Graphics,
Intel, HEFCE and PPARC.
\end{acknowledgments}



\begin{thebibliography}{}
\bibitem{reviews} V. Sahni and A. Starobinsky, Int. J.
        Mod. Phys. \textbf{D9}, 373 (2000), \texttt{astro-ph/9904398};
        T. Padmanabhan, Phys. Rept.  {\bf 380}, 235 (2003), {\tt
        hep-th/0212290}.
\bibitem{rec} D. Huterer and M. S. Turner, Phys. Rev. D{\bf 60},081301 
	(1999), {\tt astro-ph/9808133}; A. A. Starobinsky,JETP Lett. 
	 {\bf 68}, 757 (1998)[Pisma Zh. Eksp. Teor. Fiz. {\bf 68}, 721 
	 (1998)], {\tt astro-ph/9810431}; T. Nakamura and T. Chiba,
	 Mon. Not. Roy. Astron. Soc. {\bf 306}, 696 (1999), {\tt
	 astro-ph/9810447}; B. F. Gerke and G. Efstathiou, Mon. Not. 
	 Roy. Astron. Soc.  {\bf 335}, 33 (2002), {\tt astro-ph/0201336}; 
	 C. Wetterich, Phys. Lett. B{\bf 594}, 17 (2004),
	 {\tt astro-ph/0403289}.
\bibitem{GOZ} Z.-K. Guo, N. Ohta and Y.-Z. Zhang, {\tt astro-ph/0505253}.
\bibitem{simon} J. Simon, L. Verde and R. Jiminez, Phys. Rev. D{\bf 71},
	123001 (2005), {\tt astro-ph/0412269}.
\bibitem{CKLL} E. J. Copeland, E. W. Kolb, A. R. Liddle and J. E. Lidsey,
	Phys. Rev. D{\bf 48}, 2529 (1993), {\tt hep-ph/9303288}.
\bibitem{GL} I. J. Grivell and A. R. Liddle, Phys. Rev. D{\bf 61}, 081301 
	(2000), {\tt astro-ph/9906327}.
\bibitem{Riess:2004nr} A. G. Riess {\it et al.}  [Supernova Search Team 
	Collaboration], Astrophys. J.  {\bf 607}, 665 (2004),
	{\tt astro-ph/0402512}.
\bibitem{Verde:2003ey} L. Verde {\it et al.}, Astrophys.  J. Suppl. 
	{\bf 148}, 195 (2003), {\tt astro-ph/0302218}.
\bibitem{Dunkley:2004sv} J. Dunkley, M. Bucher, P. G. Ferreira, K. Moodley 
	and C. Skordis, Mon. Not. Roy. Astron. Soc. {\bf 356}, 925 (2005),
	{\tt astro-ph/0405462}.
\bibitem{Lewis:2002ah} A. Lewis and S. Bridle, Phys. Rev. D {\bf 66}, 
	103511 (2002), {\tt astro-ph/0205436}.  
\bibitem{Gilks} W. R. Gilks, S. Richardson, and D. J. Spiegelhalter (eds.),
	{\em Markov Chain Monte Carlo in Practice}, Chapman \& Hall (1996).
\bibitem{GelmanRubin} A. Gelman and D. Rubin, Statistical Science {\bf 7}, 
	457 (1992).
\bibitem{Tegmark:2003ud} M. Tegmark {\it et al.}  [SDSS Collaboration],
	Phys. Rev. D{\bf 69}, 103501 (2004),{\tt astro-ph/0310723}.
\bibitem{jeff} H. Jeffreys, {\em Theory of Probability}, 3rd ed, Oxford
        University Press (1961).
\bibitem{mackay} D. J. C. MacKay, {\em Information theory, 
        inference and learning algorithms}, Cambridge University Press 
        (2003).
\bibitem{Lid} A. R. Liddle, Mon. Not. Roy. Astron. Soc. {\bf 351}, L49 
	(2004), {\tt astro-ph/0401198}.
\bibitem{SWB} T. D. Saini, J. Weller and S. L. Bridle, Mon. Not. Roy. 
	Astron. Soc. {\bf 348}, 603 (2004), {\tt astro-ph/0305526}.
\bibitem{Schwarz} G. Schwarz, Annals of Statistics {\bf 5}, 461 (1978).
\bibitem{Muk98} S. Mukherjee, E. D. Feigelson,  G. J. Babu, F. Murtagh, C.
	Fraley, and A. Raftery, Astrophys. J. {\bf 508}, 314 (1998), 
	{\tt astro-ph/9802085}.
\bibitem{Padmanabhan:2004av} T.~Padmanabhan, Curr. Sci. {\bf 88}, 1057 
	(2005), {\tt astro-ph/0411044}.
\bibitem{Peri} L. Perivolaropoulos, Phys. Rev. D{\bf 71}, 063503 (2005), 
	{\tt astro-ph/0412308}.
\bibitem{BCK} B. A. Bassett, P. S. Corasaniti, and M. Kunz, Astrophys. 
	J. Lett. {\bf 617}, L1 (2004), {\tt astro-ph/0407364}.
\bibitem{Maor:2002rd} I. Maor and R. Brustein, Phys. Rev. D{\bf 67}, 
	103508 (2003), {\tt hep-ph/0209203}.

\end{thebibliography}
\end{document}